\documentclass[journal,twoside,web]{ieeeconf}
\usepackage{cite}
\usepackage{amsmath,amssymb,amsfonts}
\usepackage{algorithmic}
\usepackage{graphicx}
\usepackage{textcomp}

\usepackage{amsmath,amssymb,euscript ,yfonts,psfrag,latexsym,dsfont,graphicx,bbm,color,amstext,wasysym,subfig,flushend,parskip}
\usepackage{times,mathptmx,epsfig,graphicx,lineno,color,balance}
\usepackage{caption}

\graphicspath{{./},{./figures/}}

\definecolor{grey}{rgb}{0.6,0.6,0.6}
\definecolor{lightgray}{rgb}{0.97,.99,0.99}

\definecolor{llgrey}{rgb}{0.9,0.9,0.9}
\definecolor{lgrey}{rgb}{0.6,0.6,0.6}
\definecolor{lred}{rgb}{0.9,0.7,0.7}

\newcommand{\black}{\color{black}}
\definecolor{dblue}{rgb}{0,0.5,.7}

\usepackage{xcolor}
\modulolinenumbers[5]

\newtheorem{prop}{Proposition}

\newtheorem{remark}{Remark}
\newtheorem{lemma}{Lemma}

 \newcommand{\dbar}{d\hspace*{-0.1em}\bar{}\hspace*{0.2em}}
\newcommand{\trace}{{\rm trace\,}}
\newcommand{\cQ}{Q}

\newcommand{\ud}{d}
\def\spacingset#1{\def\baselinestretch{#1}\small\normalsize}
\spacingset{1}

\begin{document}

\title{Minimal entropy production in the presence of anisotropic fluctuations}

\author{Olga Movilla Miangolarra$^\star$, Amirhossein Taghvaei$^\dagger$, and Tryphon T. Georgiou$^\star$
\thanks{$^\star$Mechanical and Aerospace Engineering, University of California, Irvine, CA 92697, USA;
omovilla@uci.edu, tryphon@uci.edu}
\thanks{$^\dagger$Aeronautics and Astronautics Department, University of Washington, Seattle, Washington 98195, USA; amirtag@uw.edu}}

\maketitle

\begin{abstract}
Anisotropy in temperature, chemical potential, or ion concentration, provides the fuel that feeds dynamical processes that sustain life. 
%
At the same time, anisotropy is a root cause of incurred losses manifested as entropy production.
In this work we consider a rudimentary model of an overdamped stochastic thermodynamic system in an anisotropic temperature heat bath, and study minimum entropy production when driving the system between thermodynamic states in finite time.\\[1pt]\hspace*{7pt}
While entropy production in \emph{isotropic} temperature environments can be expressed in terms of the length (in the Wasserstein $W_2$ metric) traversed by the thermodynamic state of the system, anisotropy complicates substantially the mechanism of entropy production since, besides dissipation, seepage of energy between ambient anisotropic heat sources by way of the system dynamics is often a major contributing factor. A key result of the paper is to show that in the presence of anisotropy, minimization of entropy production can once again be expressed via a modified Optimal Mass Transport (OMT) problem. However, in contrast to the isotropic situation that leads to a classical OMT problem and a Wasserstein length, entropy production may not be identically zero when the thermodynamic state remains unchanged (unless one has control over non-conservative forces); this is due to the fact that maintaining a Non-Equilibrium Steady-State (NESS) incurs an intrinsic entropic cost that can be traced back to a seepage of heat between heat baths.\\[1pt]\hspace*{7pt}
As alluded to, NESSs represent hallmarks of life, since living matter by necessity operates far from equilibrium. Therefore, the question studied herein, to characterize minimal entropy production in anisotropic environments, appears of central importance in biological processes and on how such processes may have evolved to optimize for available usage of resources.
\end{abstract}


\section{Introduction}

Life on Earth is possible thanks to the temperature difference between the hot Sun and the cold stary sky. This difference provides ``negative entropy'' that organisms and complex biochemical processes feed upon \cite{opatrny2017life}. Indeed, the Sun provides photons at around 6000 Kelvin that are absorbed and then re-emitted back to the cosmos at about 300 Kelvin, a twenty fold decrease. The ``hard currency'' paid along the way is a positive entropy rate for the universe as a whole, while at the same time, the {\em anisotropy} in the thermal environment powers biological engines that make life possible \cite{battle2016broken,gnesotto2018broken}.

Our goal in the present work is to quantify the entropy rate  of
thermodynamic processes while operating in anisotropic temperature environments.
Specifically, we study a rudimentary model for a thermodynamic system operating
far from equilibrium, in contact with heat baths of different temperatures.
The salient feature of the arrangement is that the flux of heat  between the heat baths, mediated by the system dynamics, is responsible for maintaining the system far from equilibrium at the cost of a positive entropy rate. Our interest is in minimizing entropy production during thermodynamic transitions, a problem cast within the frame of stochastic control and stochastic thermodynamics.

Thermodynamics was born more than a century ago with the 
foundational work of Carnot and Clasius \cite{carnot1824reflections}. It has since impacted almost every corner of science: chemistry, physics, astronomy, biology... And yet, great many puzzles, rooted at the core notions of the subject such as irreversibility and the second law (see for example Loschmidt's paradox or Maxwell's demon), continued to be debated until the closing of the 20th century.
At that time, a new level of understanding started forming. The catalyst was the discovery of a number of fluctuation theorems \cite{sevick2008fluctuation} (Evans-Searles, Jarzynski, Crooks) and a framework \cite{sekimoto2010stochastic} (stochastic energetics) to study thermodynamic systems that are made up of  possibly only a few particles and/or operate fast and far from equilibrium.
To this end probability theory and stochastic control proved enabling.
Thermodynamic systems are modeled probabilistically. They interact with externally specified conditions that serve as inputs to steer the state according to specifications, while keeping a lid on 
entropy and energy budgets. Thus, this emergent subject of {\em stochastic thermodynamics} falls squarely within the purview of stochastic control.

The presentation in this work focuses on mesoscopic systems modeled by Langevin stochastic differential equations. In this context, it turns out that {\em dissipation} can be expressed 
as a quadratic cost functional that, in geometric terms, is precisely the Wasserstein length traversed by the state of the system \cite{aurell2011optimal,dechant2019thermodynamic,chen2019stochastic}.
Important new insights have been gained in recent years regarding optimal control laws that extract work while alternating contact with different heat sources in the Carnot model \cite{fu2021maximal},
quantifying natural time-constants and establishing uncertainty relations \cite{nakazato2021geometrical,van2022thermodynamic,ito2022geometric},
and balancing work with dissipation in finite-time thermodynamic transitions 
\cite{EnergyHarvestingAnisotropic2021,miangolarra2022geometry}. Within this evolving landscape the study of entropy production in the presence of temperature gradients remains largely unexplored, and constitutes the theme of the present work.


The structure and contributions of the paper are as follows:
In Section \ref{sec:stochasticthermo} we summarize the basic framework of stochastic thermodynamics in continuous space and time. In particular, we express thermodynamic quantities of interest as integral control costs. 
 In Section \ref{sec:control} we introduce thermodynamic states as probability densities and their tangents as gradient vectors. A weighted inner product  is defined on the tangent space to account for anisotropy, leading to a decomposition of velocity fields into gradient and divergence-free parts, {\em à la} Helmholtz, that constitutes the backbone of our work. 
%
%

Section \ref{sec:nonconservative} 
considers the problem of minimizing entropy production using non-conservative forces (control inputs). It is shown that the minimal entropic cost for transitioning between states can be expressed as a modified Wasserstein metric, 
extending earlier results for the isotropic thermal environment of a single heat bath and constituting our first main result.
Section~\ref{sec:constrained} considers the case where actuation takes the form of time-dependent potentials (potential forcing). Our second main result consists of a geometric decomposition of entropy production, based on the weighted inner product introduced in Section \ref{sec:control}, that generalizes previous works. In this decomposition, the contributions of the  gradient and divergence-free components of the velocity field are separated into {\em excess} and {\em housekeeping} entropy production. Further, divergence-free components are split into steady-state and dynamic contributions.
This turns out to be key in our search for optimal protocols to minimize entropy production. To this end, we characterize the set of tangent directions with vanishing {\em housekeeping entropy} production, and determine the one direction that in addition has the least {\em excess entropy} rate. Moreover, we determine protocols that minimize total entropy production {\em rate} and show that, when transitioning between product states, excess  and housekeeping entropy production are simultaneously minimized by geodesics in the weighted Wasserstein space.

Finally, Section \ref{sec:quadratic}
specializes results of the previous sections to quadratic controlling potentials, while in Section \ref{sec:2D} we give explicit expressions for the entropy production and optimizing controls for a $2$-dimensional example. 
Specifically, we show that i) nontrivial trajectories with vanishing housekeeping entropy production exist, ii) we characterize directions that minimize the production rate for the total entropy, and iii) we explicitly describe trajectories,  close to equilibrium, that minimize total entropy production. Lastly, iv) we display closed cycles that generate minimal amount of entropy, which, in contrast to the isotropic case, do not remain static and gravitate towards an equilibrium state. 
\black

\section{Stochastic Thermodynamic Systems}\label{sec:stochasticthermo}

Stochastic Thermodynamics \cite{sekimoto2010stochastic,seifert2012stochastic,peliti2021stochastic} has been successful in modeling mesoscopic thermodynamic processes that evolve both in discrete as well as in continuous state space, utilizing Master Equations or Langevin Stochastic Differential equations, respectively; herein we restrict our attention to the latter.
A thermodynamic system, at a mesoscopic scale, can be conceptualized as a collection of particles in contact with heat baths, modeled as sources of stochastic excitation, while driven under the influence of external forces. These forces that may represent control inputs can be conservative (gradients) or non-conservative, and the system dynamics may or may not include inertial effects. 


In the present work we focus on overdamped dynamics, that is, we neglect inertial effects. Such models are typical when considering colloidal mesoscopic particle systems and models of biological processes 
\cite{blickle2012realization,seifert2012stochastic,ciliberto2017experiments}. 
While conventional overdamped systems represent a collection of particles in contact with a single heat bath at any time, we consider a more general setting, where the thermodynamic system has $n$ coupled degrees of freedom which are subject to fluctuations of different intensities;  that is, the system is in contact with multiple heat baths at the same time. 
Thus, our basic model is the following Langevin system of stochastic differential equations
 \begin{equation}\label{eq:Langevin}
dX_t=-\gamma^{-1}\nabla U(t,X_t) dt+ \gamma^{-1}f(t,X_t)dt+ \sqrt{2D}dB_t, 
 \end{equation}
where $X_t\in \mathbb R^n$, $t\in\mathbb R$ represents time, $\nabla U(t,X_t)$ represents the conservative forces of the drift term (typically constituting our control), $f$ represents non-conservative forces, $B_t \in \mathbb R^n$ is a standard Brownian motion, and $D$ represents the diffusion tensor that abides by the Einstein relation
\begin{equation*}
D = k_B \gamma^{-1} T,
\end{equation*}
with 
\[
T={\rm diag}\,(T_1, T_2,\ldots, T_n),
\]
a diagonal matrix with entries the value of temperature (in Kelvin) along the specified $n$ degrees of freedom, $\gamma$ a scalar friction coefficient, and $k_B$ the Boltzmann constant\footnote{The Boltzmann constant has dimensions [energy/degree Kelvin], the same as the units of entropy as is the convention in the physics literature, cf.\ Section \ref{sec:2ndlaw}, Equation \eqref{eq:Ssys}.}.
The degrees of freedom in \eqref{eq:Langevin} may represent voltages in an electrical circuit with resistors subject to different Johnson-Nyquist thermal noise~\cite{BGyrator2017electrical}, or a single particle subject to radiation of different intensity from different directions~\cite{bang2018experimental}.


The state of the thermodynamic system is represented by the probability density function $\rho(t,x)$ that satisfies the Fokker-Planck equation\footnote{As is common, for $\partial_i:=\frac{\partial}{\partial x_i}$, $\nabla := \left(\begin{matrix}\partial_1, &\ldots, & \partial_n\end{matrix}\right)^\prime$  denotes the gradient and ``$\nabla\cdot\;\;$'' the divergence.}
\begin{equation}\label{eq:FP}
\partial_t \rho(t,x) +\nabla \cdot J(t,x) =0,
\end{equation}
with $x=(x_1,\ldots,x_n)^\prime\in\mathbb R^n$ and {\em probability current}
\begin{align}\nonumber
J(t,x)&=-\rho(t,x)\gamma^{-1}
\left(
\nabla U(t,x) - f(t,x) + k_BT\nabla \log\rho(t,x)
\right) \\&=: \rho(t,x) v(t,x). \label{eq:pcurrent}
\end{align}
Thereby, $v(t,x)$ defined above may be seen to represent the {\em velocity field} of an ensemble of particles.
 For the most part (Section \ref{sec:constrained} and on), we will assume the absence of non-conservative forces, i.e., that $f=0$.

\subsection{The first law}
During thermodynamic transitions,
energy is continuously exchanged between particles, external actuation and the surrounding heat bath  through heat and work \cite{sekimoto2010stochastic}.

The incremental change in internal energy 
$E_t=U(t,X_t)$ of a single particle at location $X_t$
can be expressed as
\begin{align*}
    dE_t&=\partial_t U(t,X_t)dt+\nabla U(t,X_t)'\circ dX_t
\end{align*}
where $\circ$ denotes Stratonovich integration.
The heat exchanged at the scale of a single particle is due to forces\footnote{Below, $\frac{dB_t}{dt}$ is formal and represents white noise and similarly for $\frac{dX_t}{dt}$, but these can also be interpreted e.g., as in \cite{hida2013white}.}
\[
-\gamma \frac{dX_t}{dt}+\sqrt{2k_B\gamma T}\frac{dB_t}{dt}
\]
applied by the heat bath (see \cite[page 19]{sekimoto1998langevin}, \cite[page 63]{peliti2021stochastic}); the 
term $-\gamma \frac{dX_t}{dt}$ is due to dissipation, while $\sqrt{2k_B\gamma T}\frac{dB_t}{dt}$ is due to fluctuations.
Thus, the energy exchange between the heat bath and the particle  is formally expressed as forces times displacement, 
\begin{align*}
    \dbar q 
    &=\Big(-\gamma \frac{dX_t}{dt} +\sqrt{2k_B\gamma T}\frac{dB_t}{dt}\Big)'\circ dX_t,
\end{align*}
which, in combination with~\eqref{eq:Langevin}, takes the precise  mathematical form
\begin{align*}
    \dbar q 
    &=(\nabla U(t,X_t)-f(t,X_t))'\circ dX_t.
\end{align*}

The incremental work on the other hand, effected through interaction of the particle with the potential $U$ and, possibly, an external nonconservative force $f$, is
\begin{equation*}\label{eq:w}
    \dbar w= \partial_t U(t,X_t)dt +f(t,X_t)'\circ  dX_t.
\end{equation*}
Direct inspection validates the first law of thermodynamics, that energy is conserved, 
\begin{align*}
    dE &= \dbar q  + \dbar w. \label{eq:1stlaw}
\end{align*}
Note that in the above, $\dbar$ designates an inexact differential, in that 
$\int \dbar w$ along a curve depends on the choice of curve and not only the endpoints. Also, note that
the sign convention is chosen such that both work and heat are positive when supplied to the particle.


The total heat differential is the sum of 
contributions from heat baths, namely, 
\begin{equation}\label{eq:dq}
\dbar q=\sum_i \dbar q_i=\sum_i (\partial_i U(t,X_t)-f_i(t,X_t))\circ (dX_t)_i.
\end{equation}
\black
Upon using the It\^o rule to express this as an It\^o differential and taking the expectation, 
the heat rate that flows into the thermodynamic \emph{system} from the $i$-th reservoir becomes\footnote{Throughout, $dx$ is a short for the volume form $dx_1\dots dx_n$.}
\begin{equation}\label{eq:ithheat}
    \dot{Q}_i=\int(\partial_iU(t,x)-f_i(t,x))J_i(t,x) dx.
\end{equation}
Here, the influx of heat from the $i$-th  reservoir takes place along the $i$-th degree of freedom, which is coupled to the rest via the potential $U$.

\subsection{The second law}\label{sec:2ndlaw}

During thermodynamic transitions, the entropy production includes two terms, entropy production within the system and entropy change in the environment,
\[
\dot S_{\rm tot} = \dot S_{\rm sys}+\dot S_{\rm env}.
\]
The entropy of the system $S_{\rm sys}=-k_B \int \log (\rho) \rho dx$ changes with the rate\footnote{Throughout, we use the notation $\langle v_1,v_2\rangle=v_1^\prime v_2$, for the standard Euclidean inner product, where ``$\phantom{\cdot}^\prime$'' denotes transpose. Also, we use the notation $\langle v_1,v_2\rangle_M=v_1^\prime Mv_2$ for weighted inner product with a symmetric matrix $M$ and, accordingly,  $\|v\|_M^2:=\langle v,v\rangle_M^{1/2}$ for the corresponding norm.},
\begin{align}
\dot S_{\rm sys}&=-k_B\int  \log(\rho) \partial_t \rho\,dx= -k_B\int \langle J, \nabla \log\rho\rangle dx \label{eq:Ssys}
\end{align}
where the second equality follows from $\partial_t \rho = -\nabla \cdot J$ and integration by parts. The entropy of the environment changes due to the heat exchange according to
\begin{align}
\dot S_{\rm env}&=-\sum_i \frac{\dot \cQ_i}{T_i}=-\int \langle J,T^{-1}(\nabla U-f)\rangle dx,\label{eq:Senv}
\end{align}
where we have used \eqref{eq:ithheat}.
The minus sign is due to positive heat rate $\dot\cQ_i$ being taken out of the environment and into the system.

Together \eqref{eq:Ssys} and \eqref{eq:Senv} give the total entropy production 
\begin{subequations}
\begin{align}\nonumber
\dot S_{\rm tot}&=-
\int \langle J, T^{-1} \left(\nabla U -f + k_BT\nabla \log\rho\right)\rangle dx\\
&= \int \frac{1}{\rho}\|J\|^{2}_{ \gamma T^{-1}}dx,
\label{eq:entropyproduction0}
\end{align}
using \eqref{eq:pcurrent}. 
Therefore, a non-vanishing probability current $J$ 
irreversibly increases the total entropy; this constitutes the second law of thermodynamics, i.e. $\Delta S_{\rm tot}\geq 0$. Alternatively, the entropy production rate expressed in terms of $v$ is 
\begin{align}
\dot S_{\rm tot}= \gamma\int \rho\|v\|^{2}_{ T^{-1}}dx.\label{eq:entropyproduction}
\end{align}
\end{subequations}

Non-vanishing probability currents arise during thermodynamic transitions, but also at certain steady-states termed {\em non-equilibrium steady-states} (NESS). At a NESS (i.e., where $\nabla \cdot J=0$ but $J\neq 0$)
the probability current mediates heat transfer between thermal baths and an increase in the entropy of the environment.
The condition $J=0$, where entropy production vanishes, is referred to as {\em detailed balance} or {\em micro-canonical reversibility}, and the state as an equilibrium steady-state.

\black

\subsection{A fluctuation theorem}\label{sec:fluctuation}

Early achievements that helped launch the subject of stochastic thermodynamics took the form of fluctuation theorems that quantified the probability of thermodynamic transitions as a function of entropy production, see e.g., \cite{sevick2008fluctuation,jarzynski2011equalities}. In a similar spirit, we herein present a fluctuation theorem that applies in the context of anisotropic environments, where the temperature $T$ is a two-tensor.

{

 Consider a single random realization $\{X_t \in \mathbb R^n;\,t\in[0,t_f]\}$,  interpreted as the trajectory of a particle that is an element of an ensemble distributed according to $\rho(t,x)$. 
Much like internal energy, heat, and work that can be defined at the level of individual particles, the entropy of the system too can be defined for an individual particle at $X_t$ within the ensemble as $-k_B\log \rho(t,X_t)$. Thus, the entropy of the system at the level of the ensemble, $-k_B\int \rho(t,x)\log \rho(t,x)dx$,  can be interpreted as the mean entropy of particles.

Therefore, the change in the entropy of the system, at the level of a single particle, is
\begin{align*}
    \Delta s_{\rm sys} &=  -k_B\log \rho(t_f,X_{t_f}) + k_B\log \rho(0,X_0),
    \\&= -\int_0^{t_f} k_B\nabla \log(\rho(t,X_t))' \circ d X_t - \int_0^{t_f}k_B \frac{\partial_t \rho(t,X_t)}{\rho(t,X_t)} dt.
\end{align*}
Similarly, the change in the entropy of the environment is
\begin{align*}
    \Delta s_{\rm env} &=  -\int_0^{t_f} \sum_{i=1}^n \frac{\dbar q_i}{T_i} \\&=  -\int_0^{t_f}  (T^{-1}\nabla U(t,X_t) - T^{-1}f(t,X_t))' \circ dX_t.
\end{align*}

\begin{prop}
The total entropy production for a single trajectory can be expressed as
\begin{align*}
   \Delta s_{\rm tot} &= \Delta s_{\rm env} +  \Delta s_{\rm sys}\\ &= \int_0^{t_f}   \gamma T^{-1} v(t,X_t)'\circ d X_t + k_B\int_0^{t_f} \frac{\nabla \cdot J(t,X_t)}{\rho(t,X_t)} d t,
\end{align*}
and satisfies the identity
\begin{align}\label{eq:FT}
    \mathbb E\left[\exp\left(-\frac{\Delta s_{\rm tot} }{k_B} \right)\right]=1.
\end{align}
\end{prop}
}     The proof of this statement is given in the Appendix, where we also prove a detailed fluctuation theorem.

\begin{remark}
   The fluctuation theorem \eqref{eq:FT} provides a stochastic description of the second law of thermodynamics, highlighting the fact that the decrease of total entropy \emph{at the level of a single trajectory} is possible, even if unlikely. One can use Jensen's inequality to derive the standard second law, $\Delta S_{\rm tot}\geq 0$, that is, the total entropy production \emph{at the level of the ensemble} cannot decrease. $\Box$
\end{remark}

\section{Thermodynamic Space and Decomposition of Vector Fields} \label{sec:control}

Thermodynamic states are represented by probability distributions on $\mathbb R^n$. Throughout they are assumed to have finite variance and to be absolutely continuous with respect to the Lebesgue measure, thus represented by density functions\footnote{The theory that follows can be developed for spaces of probability measures $\rho$~\cite[Sec.\ 8]{ambrosio2005gradient}; we specialize to probability densities for simplicity of the exposition.}.
We additionally consider the following  mild assumptions on 
thermodynamic states and corresponding vector-fields:

\noindent
{\bf Assumption A1:}
For all thermodynamic states $\rho(x)$ the following hold:
\begin{align*}
    {\rm i)}\;\; & \|\nabla^2 \log \rho\|_\infty < \infty\\
    {\rm ii)}\;\; & \|\nabla \log \rho(x)\| \to \infty \mbox{ as }\|x\| \to \infty.
\end{align*}

\noindent
{\bf Assumption A2:} Vector-fields $v(t,x)$ are Lipschitz.


Assumption A1 ensures that $\rho$ satisfies the Poincar\'e inequality\footnote{The Poincar\'e inequality essentially expresses that $0$ is an isolated eigenvalue of a corresponding Laplace operator.}~\cite{laugesen2015poisson}, i.e. that there exists a positive constant $C>0$ such that
\begin{equation*}
     \int|h-\bar h|^2\rho dx\leq C\int \|\nabla h\|^2\rho dx \ \mbox{ for all }h\in H^1_\rho,
\end{equation*}
where $\bar h=\int h\rho dx$ and $H^1_\rho$ is the Sobolev space of functions where the function and its derivative, defined in the weak sense, are square integrable with respect to $\rho$, i.e., that both $\|h\|^2_\rho:=\int h^2 \rho dx$, and
$\int \|\nabla h\|^2 \rho dx$ are bounded. The Poincar\'e inequality is of importance as it provides a sufficient condition for existence and uniqueness of the solution $\phi \in H^1_\rho$ to the (weighted) Poisson equation 
\begin{equation*}
    \mathcal L_\rho \phi = \rho h
\end{equation*}
where $\mathcal L_\rho (\cdot):=\nabla \cdot(\rho \nabla (\cdot))$ is the (weighted) Laplacian and $\|h\|_\rho <\infty $~\cite{laugesen2015poisson}. 

    

 The space of thermodynamic states $\rho$ is denoted by $P_2(\mathbb R^n)$ (or, $P_2$ for simplicity). Interestingly, this space admits a very rich structure that renders it almost a Riemannian manifold \cite[page 168]{ambrosio2005gradient}. Much of what follows to a large degree can be traced to this fact.

Our starting point is the following inner-product between vector fields on $\mathbb R^n$ 
\begin{equation}\label{eq:inner-product}
\langle v_1,v_2\rangle_{\rho}= \int \langle v_1(x),v_2(x)\rangle \rho(x) \,dx,
\end{equation}
with induced norm $\|v\|_\rho:=\sqrt{\langle v,v\rangle_\rho}$. The inner-product defines an orthogonal decomposition 
\begin{equation*}
    v= \Pi_\rho\, v +\chi,
\end{equation*}
where $\Pi_\rho$ is the projection operator given by
\[
\Pi_\rho v := \arg\min_{w}\{\|w\|_\rho \,\mid\,w=v-\chi \mbox{ and }\nabla\cdot(\rho \chi)=0\}.
\]
 The projection is unique and belongs to the closure of the space of vector fields of gradient form with respect to the $\|\cdot\|_\rho$ topology~\cite[Lemma 8.4.2]{ambrosio2005gradient}. 
Under assumptions A1 and A2, the projection is exactly  of gradient form $\nabla \phi$, where $\phi \in H^1_\rho$ solves the (weighted) Poisson equation
$
    \mathcal L_\rho \phi = \nabla \cdot (\rho v) 
$.
The decomposition $v=\nabla \phi + \chi$, into the gradient and divergence-free parts, is known as the {\em Helmholtz-Hodge decomposition}  \cite{morse1953methods,chorin1990mathematical}.

This decomposition is used to construct the tangent space at any $\rho\in P_2$. In particular, for
an admissible rate of change $\dot \rho =- \nabla \cdot (\rho v)$, induced by the vector field $v$, there is a corresponding gradient vector field $\nabla \phi = \Pi_\rho\,v$, which constitutes the tangent vector. 
The correspondence between $\dot \rho$ and  $\nabla \phi$
is used to equip $P_2$ with the Riemannian metric
 \[
 \langle \dot \rho_1,\dot \rho_2\rangle_{g_\rho}:=\langle \nabla \phi_1, \nabla \phi_2 \rangle_\rho.
 \]
 The Riemannian metric
allows computing length of paths between densities; the smallest distance (geodesic) between any two given densities $\rho_0$ and $\rho_f$ is known as the {\em Wasserstein metric} $W_2(\rho_0,\rho_f)$. This is given by
\begin{equation}\label{eq:Wasserstein}
    W_2(\rho_0,\rho_f)^2=
    \min_{\rho}\int_0^1\|\dot{\rho}\|_{g_\rho}^2dt
\end{equation}
subject to 
$\rho(0)=\rho_0$, $\rho(1)=\rho_f$.

The geometrical construction described above is generalized by replacing the inner-product~\eqref{eq:inner-product} by a weighted inner-product 
\[
\langle v_1,v_2\rangle_{\rho M}= \int \rho(x) \langle v_1(x),v_2(x)\rangle_{M} dx,
\]
where $M$ is a symmetric positive-definite $n\times n$ matrix, and $\langle v_1(x),v_2(x)\rangle_{M}:=v_1(x)^\prime Mv_2(x)$. This is important, in light of \eqref{eq:entropyproduction}, where a weighted inner-product with $M=\gamma T^{-1}$ characterizes the entropy production.  In a similar manner as before, vector fields can be decomposed into gradient and divergence-free parts,
\begin{equation}\label{eq:orthogonal}
v=M^{-1}\nabla \phi + \chi,
\end{equation}
where $M^{-1} \nabla \phi = \Pi_{\rho M} v$ is now the projection with respect to the weighted metric $\langle\cdot,\cdot\rangle_{\rho M}$, and $\nabla\cdot 
(\rho\chi)=0$.
Analogously, we introduce the Riemannian metric 
\[
\langle \dot \rho_1,\dot \rho_2\rangle_{g_{\rho M}}:=\langle M^{-1}\nabla \phi_1,M^{-1}\nabla \phi_2\rangle_{\rho M},
\]
and define the weighted Wasserstein metric
\begin{align}\label{eq:W2M}
    W_{2,M}^2(\rho_0,\rho_{f})= \int_0^{1}  \|\dot \rho \|^2_{g_{\rho M}} dt
\end{align}
subject to 
$\rho(0)=\rho_0$, $\rho(1)=\rho_f$.

\section{Non-conservative actuation}\label{sec:nonconservative}

We start by considering the problem of minimizing entropy production under the full authority of non-conservative actuation, that is, with control actuation that entails both a gradient $\nabla U$ of a potential as well as a non-zero term $f$ in \eqref{eq:Langevin} contributing with a divergence-free component. This amounts to full control authority over the velocity vector-field $v$. 
It turns out that the minimal entropy production relates to a suitably weighted Wasserstein distance between states. This result extends 
the geometric characterization of entropy production in \cite{aurell2011optimal} to the case of anisotropic thermal environment. 

\subsection{Entropy production as a weighted Wasserstein length}
The entropy production can be expressed in terms of the  weighted Wasserstein distance between states as follows.
\begin{prop} It holds that
  \begin{align}\label{eq:dynamicW2}
     \min_{\rho,v}\int^{t_f}_0\dot S_{\rm tot}\, dt  = \frac{1}{t_f}W_{2,M}^2(\rho_0,\rho_{f})\end{align}
where $M=\gamma T^{-1}$ and the optimization is subject to the continuity equation $\partial_t\rho+\nabla\cdot (\rho v)=0$ together with the end-points $\rho_0$, $\rho_f$ of a path $\rho(t,\cdot)$, $t\in[0,t_f]$, effected via a control that includes gradient $\nabla U$ as well as divergence free $f$ component.
\end{prop}
\black
\begin{proof}
It readily follows by comparing the expression for the least entropy production \eqref{eq:entropyproduction} over paths $\rho(t,\cdot)$, $t\in[0,t_f]$, between end-point states, with the definition of the weighted Wasserstein metric \eqref{eq:W2M}.
\end{proof}

\begin{remark}
  If a bound is imposed on the total entropy production $S_{\rm tot}$, then  \eqref{eq:dynamicW2} provides a lower bound (speed limit) on the time needed for traversing a path that joins $\rho_0$ to $\rho_f$, namely,
$$
t_f\geq \frac{W_{2,M}^2(\rho_0,\rho_{f})}{S_{\rm tot}}.\ \Box
$$
\end{remark}


For computational purposes it is useful to relate the weighted Wasserstein distance $W_{2,M}$ to an un-weighted (corresponding to the identity matrix as weight) Wasserstein distance. Thereby, the entropy production can likewise be expressed in terms of the (unweighted) Wasserstein length. This is given below.

\begin{prop}\label{prop:prop1}
  It holds that
\begin{align*}
    \min_{\rho,v}\int^{t_f}_0\dot S_{\rm tot}\, dt  =\frac{1}{t_f} \frac{\gamma}{\sqrt[n]{\det(T)}}W_2^2( \mathbf{T}^{-\frac{1}{2}}\# \rho_0, \mathbf{T}^{-\frac{1}{2}}\# \rho_f),
\end{align*}
where $\mathbf T=T/\sqrt[n]{\det(T)}$ is volume preserving and the optimization is subject to the continuity equation $\partial_t\rho+\nabla\cdot (\rho v)=0$ together with the end-point conditions $\rho(0)=\rho_0$ and $\rho(t_f)=\rho_f$.
\end{prop}

\begin{proof}
The statement follows immediately after we express the weighted Wasserstein distance in terms of the ordinary (unweighted) distance.
To this end, we first invoke the fact that an optimal transportation plan requires constancy of the velocity along paths (in Lagrangian view point); this is standard and follows using the Cauchy-Schwartz
inequality. Thus, for a mass element (particle) that starts at location $x$ and terminates at $y$ over the time interval $[0,t_f]$, the optimal velocity remains constant 
 and equal to \begin{align*}\label{eq:v}
    v(X(x,t),t)=(y-x)/t_f
\end{align*}
with the path traversed by the particular particle being the line segment $X(x,t)=x+tv$ for $t\in[0,t_f]$.
This well-known fact
turns the dynamic optimal transport \eqref{eq:W2M} into a static (Kantorovich-type) problem, so as to be subsequently cast as an unweighted transport problem via a change of variables, as follows. 

Let
$\pi$ be a distribution on the product space $(x,y)\in\mathbb R^n\times \mathbb R^n$ that represents the law of pairing origin $x$ to destination $y$, under a transport policy. Thus $\pi$ is a coupling of random variables $X(x,0)$ and $X(y,t_f)$, with probability density functions $\rho_0(x)$ and $\rho_f(y)$, respectively; these are marginal distributions of $\pi$ and this is the only condition for $\pi$ to be a ``coupling.'' Then,
\begin{align*}
    W_{2,M}^2(\rho_0,\rho_f) &= \min_{\pi} \int \|x-y\|_M^2 d \pi 
\\&= 
\min_{\pi} \int \|M^{\frac{1}{2}}x-M^{\frac{1}{2}}y\|^2 d \pi 
\\&=W_2^2( M^{\frac{1}{2}}\# \rho_0, M^{\frac{1}{2}}\# \rho_f)
\end{align*}
where, with a slight abuse of notation, $M^{\frac{1}{2}}\# \rho_0$ denotes the push-forward
with the map\footnote{
The density of the push-forward $\rho_1=g\sharp \rho_0$ for a differentiable $g:x\mapsto y=g(x)$ is
$\rho_1(y)= \sum_{x\in g^{-1}(y)} \frac{\rho_0(x)}{|\det(\partial g/\partial x)|}$.} $g:x\mapsto M^{\frac{1}{2}}x$.
Using standard theory \cite{villani2003topics}, the optimal transport map for unweighted transport is given by the gradient of a convex function $\varphi$, and hence we now have $x\mapsto y =M^{-\frac{1}{2}} \nabla \varphi(M^{\frac{1}{2}} x)$; here, $\nabla \varphi$ is the optimal transport map between $M^{\frac{1}{2}}\# \rho_0$ and $M^{\frac{1}{2}}\# \rho_f$ for unweighted cost. In light of \eqref{eq:dynamicW2}, we arrive at the claimed expression.
\end{proof}

 A geometrical procedure to find the optimal transportation is to "warp" the space according to  $\mathbf T^{-1/2}$, identify the optimal transport in the usual way, and then "warp"  back.

\subsection{Dissipation for Gaussian thermodynamic states.}

In general, for standard optimal mass transport problems, explicit solutions are hard to come by and need to be computed numerically.
One exception is the case where the  transport traces paths on the submanifold of Gaussian distributions driven by a quadratic potential. In such cases the Wasserstein distance can be written down explicitly.

For later reference, we provide here the expression for the weighted Wasserstein-2 distance between two normal distributions:
\begin{align*}W_{2,M}(\rho_0,\rho_f)=\Big[&\|\mu_0-\mu_f\|^2_M+ \trace \big\{\Sigma_0 M+\Sigma_fM\\&{-2(\Sigma_f^{1/2}M\Sigma_0 M\Sigma_f^{1/2})^{1/2}
}\big\}\Big]^{1/2},
\end{align*}
where $\rho_0=\mathcal N(\mu_0,\Sigma_0)$ and $\rho_f=\mathcal N(\mu_f,\Sigma_f)$ are Gaussian with mean $\mu_0$ and $\mu_f$, and covariance $\Sigma_0$ and $\Sigma_{f},$ respectively.
{The corresponding optimal trajectory (displacement interpolation) is $\{\rho(t)=\mathcal N(\mu_t,\Sigma(t)) \mid t\in[0,t_f]\}$, where
\begin{subequations}\label{eq:displacement-interpol}
\begin{align}
\mu_t&=\mu_0+\frac{t}{t_f}(\mu_f-\mu_0),\\
    \Sigma(t)&=\bigg(\Big(1-\frac{t}{t_f}\Big) {\rm Id} +\frac{t}{t_f}A\bigg)\Sigma_0\bigg(\Big(1-\frac{t}{t_f}\Big) {\rm Id} +\frac{t}{t_f}A\bigg)',\label{eq:displacement-interpolb}
    \end{align}
\end{subequations}
with $A=\Sigma_f^{1/2}\big(\Sigma_f^{1/2}M\Sigma_{0}M\Sigma_f^{1/2}\big)^{-1/2}\Sigma_f^{1/2}M$, and ${\rm Id}$ is the identity matrix.}
One can derive these results from the standard (unweighted) Gaussian expressions, as explained in Proposition \ref{prop:prop1}.

\section{
Conservative actuation} \label{sec:constrained}

Herein we consider entropy production 
when the control actuation is limited to conservative forces, i.e., $f=0$ and the forcing is exerted solely by varying the potential $U$. This is physically more meaningful and easier to realize experimentally.

We note that even if the control is so constrained, it is possible to steer the thermodynamic system between arbitrary thermodynamic states
by controlling the potential function $U$. To see this, note that a selection of $\nabla U$
can specify the gradient part of 
\begin{align}\label{eq:vel-conservative}
v&=-(\gamma^{-1}\nabla U+D\nabla \log \rho),
\end{align}
 and therefore, any value for $\partial_t\rho$ in \eqref{eq:FP}.
Specifically, for any $\partial_t\rho=\dot\rho$ with $\|\frac{\dot \rho}{\rho}\|^2_\rho<\infty$, the equation
\[
\nabla \cdot (\rho \gamma^{-1}(\nabla U + k_B T \nabla \log \rho
))=\dot\rho
\]
is a Poisson equation and has a (unique) solution for $U$. 

\black
%
%
\black


\subsection{Geometric decomposition of entropy production}\label{sec:geometric}

We now study the source of entropy production by identifying the contribution of the orthogonal components of the velocity field that drives the thermodynamic states. This will prove helpful in identifying optimal protocols in what follows.

We start by considering a trajectory $\rho(t,\cdot)\in P_2$ that connects end-point states $\rho_0,\rho_f$, and we let $\dot\rho:=\partial_t\rho$. According to the discussion presented in Section~\ref{sec:control}, any vector-field $v$ that realizes the trajectory, i.e.  that $\nabla \cdot(\rho v) + \dot\rho =0$, admits an orthogonal decomposition, with respect to the metric $\langle \cdot,\cdot\rangle_{\rho T^{-1}}$, as
\begin{equation}\label{eq:v-decomp}
    v = T \nabla \phi + \chi,
\end{equation}
where the gradient part 
$T\nabla \phi$ is the projection $\Pi_{\rho T^{-1}} v$ and  $\nabla \cdot(\rho \chi)=0$.  
%
The orthogonal decomposition implies  that
the entropy production $\gamma\int_0^{t_f}\|v\|_{ \rho T^{-1}}^2 dt$ can be expressed as
\begin{equation}\label{eq:decomp}
    \gamma \int_0^{t_f}\|T\nabla \phi\|_{ \rho T^{-1}}^2 dt +\gamma
\int_0^{t_f} \|\chi\|_{\rho T^{-1}}^2 dt.
\end{equation}
\black

The first of these two contributions represents the minimal entropy production that is attainable when we allow non-conservative actuation to drive the system over the specified time interval between the two states (cf.\ Section \ref{sec:nonconservative}) -- it is precisely the Wasserstein action integral for the space equipped with the $\langle \cdot,\cdot \rangle_{g_\rho T^{-1}}$  Riemannian metric.  Thus, it constitutes a lower bound to the total entropy production \eqref{eq:decomp}. It can be thought of as the entropic cost related to driving the thermodynamic system between the two states in finite time and will be denoted by $S_{\rm ex}$ (for {\em excess} cost).

The second term in \eqref{eq:decomp}
represents a contribution to the entropy production that is due to circulation in the velocity field. Such circulation, for instance, is needed to sustain a non-equilibrium steady-state (NESS). Thus, under conservative actuation, minimal entropy production can no longer be expressed in terms of a distance between end-point states, since maintaining a stationary state incurs in positive entropy production. Contribution to entropy production due to circulatory currents that are generated when steering a system out of equilibrium by  non-conservative forcing in a uniform heat bath \cite{nakazato2021geometrical,Dechant2021geometric,ito2022geometric} 
has been referred to as ``housekeeping entropy production.'' We will follow a similar convention and label the second term in \eqref{eq:decomp} as {\em housekeeping}, and denote it by $S_{\rm hk}$.

Thus, the decomposition of entropy production in \eqref{eq:decomp} can be seen as a generalization to anisotropic temperature fields of analogous decompositions discussed in  earlier works \cite{nakazato2021geometrical,Dechant2021geometric,ito2022geometric,geomdecompcoupling2022dechant}; these works consider non-conservative forcing and heat bath with uniform temperature (a single heat bath, with $T$ scalar).  At the time the present work was being completed, Yoshimura etal.\  \cite{artemydechant2023} proposed a decomposition of entropy production into housekeeping and excess entropy terms for applications to chemical reactions that is analogous to the one presented herein, albeit developed in \cite{artemydechant2023} for discrete spaces of chemical reactants.


The nature of $S_{\rm hk}$ when the thermodynamic system is steered in the presence of thermal anisotropy is considerably more involved than when sustaining a NESS\footnote{When maintaining a NESS $\rho$ with two degrees of freedom  (subject to ``hot'' and ``cold'' thermal excitation, respectively), $S_{\rm hk}=Q(\frac{1}{T_c}-\frac{1}{T_h})$, for $Q$ the heat that flows from the hot  to the cold heat bath, at temperatures $T_h$ and $T_c$, respectively.}.
While a significant component is due to leakage of heat between the heat baths by way of the coupling between the degrees of freedom, the dynamics of the system also mediate such leakage.

We make this more precise by expressing the circulation as summation of circulation due to transition  and the circulation necessary to maintain a NESS. The definition of the velocity field \eqref{eq:vel-conservative} and its orthogonal decomposition \eqref{eq:v-decomp} imply the relationship
\begin{align}\label{eq:vel-decomp2}
    v &= -\gamma^{-1}\nabla U - D \nabla \log \rho =T \nabla \phi + \chi.
\end{align}
Consider a steady-state, for which $T\nabla \phi=0$, and let $U_{\rm ss}$ and $\chi_{\rm ss}$ denote the potential function and circulation at steady-state. Then, we have the identity 
\begin{align*}
    - D\nabla \log \rho =  \gamma^{-1}\nabla U_{\rm ss} + \chi_{\rm ss},
\end{align*}
implying that $U_{\rm ss}$ and $\chi_{\rm ss}$ are the terms in the Helmholtz-Hodge decomposition of $-D\nabla \log \rho$, i.e. that
\begin{align*}
    \chi_{\rm ss} &= (\Pi_\rho - \text{Id}) D\nabla \log \rho ,\\
    \nabla U_{\rm ss}&= -\gamma \Pi_\rho D\nabla \log \rho.
\end{align*}
In general, when $T\nabla \phi \neq 0$, the potential function and circulation have additional terms; they are given by 
\begin{subequations}\label{eq:circulation-decomp}
    \begin{align}
    \chi &= \chi_{\rm ss} + (\Pi_\rho - \text{Id}) T\nabla \phi, \\
    \nabla U &= \nabla U_{\rm ss} - \gamma \Pi_\rho T\nabla \phi.  
\end{align}
\end{subequations}

This concludes the decomposition of the circulation to contributions from transitioning and maintaining a steady-state. The following lemma states that the circulation at steady-state is zero (implying equilibrium) if and only if 
all degrees of freedom are independent. 

\begin{lemma}\label{lemma}
Let $T_i\neq T_j$ for all $i\neq j$. The entropy production rate for sustaining steady-state at $\rho$ vanishes if and only if
\begin{align}\label{eq:indep}
   \rho(x)= \prod_{i=1}^n \rho_i( x_i),
\end{align}
that is, all degrees of freedom  are independent. 
\end{lemma}

\begin{proof}
The entropy production is zero at steady-state iff $\chi_{\rm ss}=0$, implying $D\nabla \log \rho$ is of gradient form. As a result, the orthogonality condition,
\begin{equation*}
    \langle T\nabla \log \rho,\chi \rangle_\rho = 0,\quad \forall \chi\quad \text{s.t.}\quad \nabla \cdot(\rho \chi)=0,
\end{equation*}
holds. 
Let $\chi= \frac{1}{\rho}\Omega \nabla \psi$ for arbitrary skew-symmetric matrix $\Omega$ and function $\psi$.  The divergence-free condition is satisfied because
\begin{align*}
    \nabla \cdot(\rho \chi) &= \nabla \cdot(\Omega \nabla \psi) =\sum_{i,j=1}^n \partial_i(\Omega_{ij}\partial_{j} \psi) \\&= \frac{1}{2} \sum_{i,j=1}^n (\Omega_{ij} + \Omega_{ji})\partial_{ij} \psi = 0.
\end{align*}
The orthogonality condition implies
\begin{align*}
    \int (\nabla \log \rho)'T\Omega \nabla \psi dx= -\int \trace(\Omega T\nabla^2 \log \rho ) \psi dx = 0
\end{align*}
Requiring this identity to hold for any choice of $\psi$ and $\Omega$ concludes $T\nabla^2 \log \rho$ to be symmetric, which is true iff $ \partial_{ij} \log(\rho) = 0$. This implies $\log(\rho(x)) = \sum_{i=1}^n \log \rho_i(x_i)$  concluding that the distribution is of product form and all degrees of freedom are mutually independent. Conversely, if $\rho$ is of product form, then $T\nabla \log \rho(x) = \nabla (\sum_{i=1}^n T_i \log \rho_i(x_i))$, thus orthogonal to any divergence free vector-field $\chi$.   
\end{proof}

\begin{remark}\label{remark:factorization}
The statement of the lemma can be easily extended to the case where some degrees of freedom correspond to the same temperature. In that case, $\rho(x)$ factors as $\prod_{i=1}^m \rho_i(\tilde x_i)$ where $m$ is the number of different temperatures and $\tilde x_i$ is the collection of degrees of freedom corresponding to $T_i$. $\Box$
\end{remark}

\subsection{Directions of vanishing housekeeping entropy production}
Unlike the steady-state circulation which is zero only if the distribution is of product form, the total circulation can vanish by steering the system in specific directions. 
These directions result in a dynamical contribution to $\chi$ that cancels out the steady-state circulation $\chi_{ss}$ (see \eqref{eq:circulation-decomp}).
In fact, there are infinitely many such directions where the circulation vanishes. We characterize these next.


\begin{prop}\label{prop:0hk}
A choice of potential $U(x)$ results in zero housekeeping entropy production (i.e., such that $\chi=0$) if and only if $\nabla U$ lies in 
the range of 
$\Pi_{\rho T^{-1}}$.
%
Specifically, if $T_i\neq T_j$ for all $i\neq j$,
\begin{align}
    \nabla U = -\gamma T \nabla \psi, 
\end{align}
where $\psi$ is any function of the form $\psi(x) = \sum_{i=1}^n\psi_i( x_i)$. 
\end{prop}
\begin{proof}
Setting $\chi=0$ in \eqref{eq:vel-decomp2} gives that
\begin{align*}
   \gamma^{-1}\nabla U= - D \nabla \log \rho -T \nabla \phi.
\end{align*}
Since both $(\Pi_{\rho T^{-1}}-{\rm Id})D \nabla \log \rho$ and $(\Pi_{\rho T^{-1}}-{\rm Id})T\nabla \phi$ are zero,
 $\gamma^{-1}\nabla U=-T \nabla \psi$ for a suitable $\psi$. It follows that if $T_i\neq T_j$ for all $i\neq j$, $\psi$ must be such that $T\nabla \psi = \nabla \psi_T$ where $\psi_T(x) = \sum_{i=1}^n  T_i \psi_i(x_i)$.
\end{proof}

A potential $U$ as in the proposition leads to tangent directions of the form \begin{align}\label{eq:vanishinghk-tangent} 
     T \nabla \phi = - D \nabla \log \rho  + T \nabla \psi,
\end{align}
having vanishing housekeeping entropy production.
Among such directions, it is interesting to characterize the one with minimum excess entropy production:
\begin{align}\label{eq:min-S-ex}
     \min \,\dot S_{\rm ex} = \min_{\nabla \phi}\,\gamma \|T\nabla \phi\|_{\rho T^{-1}}^2,\quad \text{s.t.}\quad \chi=0. 
\end{align}
This is the content of the following proposition.

\begin{prop}\label{prop:min-ex-0hk}
Let $T_i\neq T_j$ for all $i\neq j$. The minimum excess entropy production \emph{rate} with zero circulation,
\begin{align}\label{eq:ex-0hk}
    \gamma^{-1} k_B^2 \| \nabla \log \rho - \nabla \log \bar \rho \|_{\rho T}^2,
\end{align}
is attained using $\nabla U=-\gamma T\nabla\psi$ with 
\begin{equation}\label{eq:psi0hkex}
    \psi=\gamma^{-1}k_B\sum_{i=1}^n\log\rho_i(x_i).
\end{equation}
Here, $\rho_i(x_i) = \int \rho(x) dx_{/i}$, with $dx_{/i}$ denoting integration with respect to all coordinates except $x_i$, is the $i$-th marginal of $\rho$, and
$\bar \rho(x) = \prod_{i=1}^n \rho_i(x_i)$.
\black
\end{prop}
\begin{proof}
The formula for tangent directions with vanishing circulation \eqref{eq:vanishinghk-tangent} implies that
\begin{align*}
    \gamma\|T\nabla \phi \|_{\rho T^{-1}}^2 &=  \gamma \|\gamma^{-1} k_B T\nabla\log \rho - T \nabla \psi \|_{\rho T^{-1}}^2\\
    &=  \gamma^{-1} k_B^2  \|\nabla\log \rho -  \nabla  \tilde \psi \|_{\rho T}^2,
\end{align*}
with $\tilde \psi = k_B^{-1}\gamma \psi$. Minimizing the squared-norm 
\begin{align*}
     \|\nabla\log \rho -  \nabla  \tilde \psi \|_{\rho T}^2=\sum_{i=1}^n T_i \int(\partial_i \log \rho(x) - \partial_i \tilde \psi_i(x_i))^2\, \rho(x) dx,
\end{align*}
involves $n$ independent minimizations, each over  $\partial_i \tilde \psi_i(x_i)$ for $i=1,2,\ldots,n$. The solution to each minimization problem is the conditional expectation 
\begin{align*}
    \partial_i \tilde\psi_i(x_i)=\!\int \partial_i \log \rho (x) \rho(x_{/i}|x_i)\,dx_{/i},
\end{align*}
 where $\rho(x_{/i}|x_i)$ is the conditional density of $x_{/i}$ given $x_i$. Upon using the definition of conditional density $\rho(x_{/i}|x_i)=\rho(x)/\rho_i(x_i)$, 
\begin{align*}
    \partial_i \tilde\psi_i(x_i)&=\!\int \frac{\partial_i \rho (x)}{\rho(x)} \frac{\rho(x)}{\rho_i(x_i)}\,dx_{/i}= \frac{1}{\rho_i(x_i)}\partial_i \int \rho(x)d x_{/i}\\
    &= \partial_i \log \rho_i(x_i),
\end{align*}
concluding the result.
\end{proof}

\begin{remark}
Vanishing minimum excess entropy production in~\eqref{eq:ex-0hk} is achieved if and only if $\rho$ is of product form, i.e., as in Lemma \ref{lemma}. This stands to reason, since $\chi_{ss}\neq 0$ implies that the only way to make $\chi$ vanish in \eqref{eq:circulation-decomp} is through a non-zero velocity $T\nabla\phi$, leading to non-zero excess entropy. A similar remark applies to the statement of the following subsection. $\Box$
\end{remark}

\subsection{Direction of least entropy production}

Leaving trajectories with vanishing housekeeping entropy production aside, directions minimizing total entropy production rate can be identified. These are of interest, not only for  standard energetic considerations, but also because gradient flows that minimize entropy rate are envisioned to have physical  and biological significance.
Thus, below we identify the potential $U$ that minimizes instantaneous entropy production \emph{rate}. 

\begin{prop}
The vector-field $\nabla U$ that minimizes the entropy production \emph{rate}~\eqref{eq:entropyproduction} is given in terms of the orthogonal projection of $-k_B\nabla \log(\rho)$ 
with respect to $\langle\cdot,\cdot\rangle_{\rho T}$ onto gradient fields  (cf.\ \eqref{eq:orthogonal}), i.e.,
\begin{equation}
    \nabla U = -k_B T \Pi_{\rho T}(\nabla \log(\rho)).
\end{equation}
The potential function $U\in H^1(\rho)$ 
is the unique solution to the Poisson equation
\begin{equation}\label{eq:tomatch1}
    \mathcal L_{\rho T^{-1}} U= - k_B\Delta \rho,
\end{equation}
where $\mathcal L_{\rho M}(\cdot):=\nabla\cdot ( \rho M\nabla (\cdot))$.
\end{prop}
\begin{proof}
Using the decomposition
\[k_B\nabla \log \rho = T^{-1}\nabla \psi + \chi, \]
where $T^{-1}\nabla \psi = k_B \Pi_{\rho T} \nabla \log \rho$ and $\nabla \cdot(\rho \chi)=0$, the entropy production rate
\begin{align*}
    \gamma \dot S_{\rm tot} &= \|\nabla U + k_BT \nabla \log \rho\|_{\rho T^{-1}}^2 \\&= \|T^{-1}\nabla U + k_B\nabla \log \rho\|^2_{\rho T}\\&= \| T^{-1} \nabla U + T^{-1} \nabla \psi\|_{\rho T}^2 + \|\chi\|_{\rho T}^2
\end{align*}
is minimized by $\nabla U =- \nabla \psi= - k_B T \Pi_{\rho T} \nabla \log \rho$. The Poisson equation follows from the definition of the projection and the fact that $\nabla \cdot(\rho \chi)=0$.
Existence and uniqueness of a (weak) solution to the Poisson equation follows from the Poincar\'e inequality and finiteness of $\|\frac{\Delta\rho}{\rho}\|_\rho^2$, which hold under Assumption A1~\cite{laugesen2015poisson}. 
 \end{proof}

\subsection{Minimal entropy production between end-points}

We now consider minimizing entropy production along \emph{a path} between two distributions driven by conservative actuation.

We first rewrite the rate of entropy production as
\begin{align*}
   \dot S_{\rm tot} =&\gamma^{-1} \int \|\nabla U+k_BT\nabla \log \rho\|^2_{ T^{-1}}\rho dx\\=& \gamma^{-1} \Big[\int \|\nabla U\|^2_{T^{-1}} \rho dx-k_B^2 \int \|\nabla \log \rho\|^2_T\rho dx\Big]+2\dot S_{\rm sys},
\end{align*}
using that $\dot S_{\rm sys}=-k_B\int \partial_t\rho\log\rho dx=-k_B\int (\nabla\log\rho)'v\rho dx$ via integration by parts. Since $\int_0^{t_f}\dot S_{\rm sys}dt=S_{\rm sys}(\rho(t_f))-S_{\rm sys}(\rho(0))$ only depends on the end-point distributions, minimizing entropy production over the transition amounts to solving
\begin{equation}\label{eq:equiv-cost}
    \min_{U,\rho} \gamma^{-1} \int_0^{t_f} \int\Big[ \|\nabla U\|^2_{ T^{-1}} -k_B^2\|\nabla \log \rho\|^2_T\Big]\rho dxdt,
\end{equation}
subject to the continuity equation \eqref{eq:FP} and the end-point conditions. Necessary conditions for optimality are stated next, expressed as equations (\ref{eq:FONCa}-\ref{eq:FONCc}).

\begin{prop} A path $\rho(t,\cdot)$  between specified terminal states, along with the corresponding control protocol $U(t,\cdot)$ that solve
\begin{align*}
\min_{U,\rho} \Big\{  \int_{0}^{t_f}\dot S_{\rm tot} dt
    \mid \eqref{eq:FP} \mbox{ and } \rho(0)=&\rho_0,\ \rho(t_f)=\rho_f\Big\},
\end{align*}
satisfies the equations 
\begin{subequations}\label{eq:FONC}
\begin{align}\label{eq:FONCa}
\gamma \partial_t\rho  =&\nabla\cdot(\rho(\nabla U+k_BT\nabla\log\rho)),\\ \nonumber
\gamma\partial_t\lambda=&\|\nabla U\|^2_{T^{-1}} 
  +\tfrac{2k_B^2}{\rho}\nabla\cdot(T \nabla \rho)
 -k_B^2\|\nabla \log \rho\|^2_T
   \\\label{eq:FONCb}
&+\langle\nabla\lambda,\nabla U+ k_BT\nabla \log \rho\rangle
  -\tfrac{1}{\rho}\nabla\cdot(\rho k_BT\nabla\lambda),\\
  0=&\mathcal L_{\rho T^{-1}} U  +\frac{1}{2}\nabla \cdot(\rho\nabla \lambda )
  \label{eq:FONCc}.
\end{align}
\end{subequations}
with boundary values $\rho(0)=\rho_0$ and $\rho(t_f)=\rho_f$.  
\end{prop}

\begin{proof}
We use the expression in \eqref{eq:equiv-cost} to write the following augmented Lagrangian:
\begin{align*}
    J=&\gamma^{-1} \int_0^{t_f} \int\Big[ (\nabla U)^\prime T^{-1} \nabla U-k_B^2(\nabla \log \rho)^\prime T \nabla \log \rho\Big]\rho dx dt\\&+\int_0^{t_f} \int\lambda\Big[\partial_t\rho-\gamma^{-1}\nabla\cdot((\nabla U+k_BT\nabla\log\rho)\rho)\Big]dxdt,
\end{align*}
with $\lambda$ a Lagrange multiplier. The first variation is
\begin{align*}
    \delta J=&\gamma^{-1}\int_0^{t_f} \int\Big[ 2(\nabla \delta_U)^\prime T^{-1} \nabla U\rho+(\nabla U)^\prime T^{-1} \nabla U\delta_\rho\\
    &-2k_B^2\big(\nabla \tfrac{\delta_\rho}{\rho}\big)^\prime T \nabla \log \rho\rho-k_B^2(\nabla \log \rho)^\prime T \nabla \log \rho\delta_\rho\\
    &+\lambda\big\{\gamma\partial_t\delta_\rho-\nabla\cdot((\nabla U+k_BT\nabla\log\rho)\delta_\rho)\\
    &-\nabla\cdot((\nabla \delta_U+k_BT\nabla\tfrac{\delta_\rho}{\rho})\rho)\big\}\\
    &+\delta_\lambda\big(\gamma\partial_t\rho-\nabla\cdot((\nabla U+k_BT\nabla\log\rho)\rho)\big)\Big] dxdt.
\end{align*}
Integrating by parts and setting this to zero for all perturbations $\delta_U,\delta_\rho,\delta_\lambda$ we obtain \eqref{eq:FONC}.
\end{proof}

The equations \eqref{eq:FONC} have the structure of a coupled system of partial differential equations and, in general, need to be solved numerically. However, closed-form solutions to this problem can be obtained for the following special case.


\black

\subsection{Minimal entropy production between product states}\label{sec:prod-states}

We herein focus on minimizing entropy production while transitioning between two states with mutually independent degrees of freedom, i.e. 
\begin{align}\label{eq:min-entropy-prod-states}
\min_{U,\rho} \Big\{ \int_{0}^{t_f} \dot S_{\rm tot} dt
    \mid \eqref{eq:FP},\ \rho(0)=\prod_{i=1}^n \rho^0_i( x_i),\  \rho(t_f)=\prod_{i=1}^n \rho^f_i( x_i)\Big\}.
\end{align}
It turns out that this is equivalent to minimizing excess entropy production, and that the thermodynamic state retains independence along degrees of freedom, resulting in the following statement.
\begin{prop}
The solution to \eqref{eq:min-entropy-prod-states} coincides with the solution to the unconstrained problem \eqref{eq:dynamicW2} with end-points $\rho(0)$ and $\rho(t_f)$ as above.
\end{prop}

\begin{proof}
    From Section \ref{sec:geometric} we know that the minimal excess entropy production $S_{\rm ex}$, when transitioning between $\rho_0=\prod_{i=1}^n \rho^0_i( x_i)$ and $\rho_f=\prod_{i=1}^n \rho^f_i( x_i)$, is
    \begin{align*}
     \min_{\rho,v}\int^{t_f}_0\dot S_{\rm ex}\, dt  = \frac{1}{t_f}W_{2,M}^2(\rho_0,\rho_{f}),
     \end{align*}
     for $M=\gamma T^{-1}$. 
     We have that
     \begin{align*}
    W_{2,M}^2(\rho_0,\rho_f) &= \min_{\pi} \int \|x-y\|_M^2 d \pi (x,y),
  \end{align*}
  where $\pi(x_1,x_2,\ldots,y_1,y_2,\ldots)$ is a coupling having marginals $\rho_i^0(x_i)$ and $\rho_i^f(y_i)$, with $x_i,y_i$ denoting coordinates along the same $i$-th degree of freedom at the start and end of the transition, respectively. We note that the cost $\|x-y\|_M^2$ splits as $\sum_i^n \|x_i-y_i\|_M^2$ leading to $n$ uncoupled weighted OMT problems,
  \begin{align*}W_{2,M}^2(\rho_0,\rho_f) &= 
\sum_i^n \min_{\pi_i} \int \|x_i-y_i\|_M^2 d \pi_i(x_i,y_i), 
\end{align*}
 where now $\pi_i$ is a coupling between the starting and ending marginals of the $i$-th degree of freedom $\rho_i^0$ and $\rho_i^f$. Thus, the optimal velocity field along the $i$-th degree of freedom is of the form $T_i\partial_i\phi_i(x_i)$, concluding in an optimal velocity field for the original problem of the form $T\nabla \phi = \nabla \phi_T$ where $\phi_T(x) = \sum_{i=1}^n  T_i \phi_i(x_i)$. 
 Moreover, the resulting path of thermodynamic states $\rho(t,x)$ retains the product structure of the starting and ending marginals. This path can be realized by a control of the form $\nabla U=-\gamma T\nabla \psi$ as in Proposition \ref{prop:0hk}, resulting in vanishing housekeeping entropy production. 
Therefore, we have shown that the optimal protocol for minimizing excess entropy also minimizes housekeeping entropy, as claimed.
\end{proof}

\black
\section{Control via a quadratic potential}\label{sec:quadratic}
We now specialize to the case where the controlling potential is quadratic, namely,
$
U(t,x)= x^\prime  K(t) x/2
$
with $x\in \mathbb R^n$ and $K(t)=K(t)^\prime$. Starting from an initial zero-mean Gaussian distribution, the thermodynamic state traces a path on the submanifold of Gaussian distributions
\begin{equation}\label{eq:G-pdf}
\rho(t,x)=
\frac{1}{(2\pi)^{n/2}\det(\Sigma(t))^{1/2}} e^{-\frac12 \|x\|^2_{\Sigma(t)^{-1}}},
\end{equation}
where the covariance $\Sigma$ satisfies the differential Lyapunov equation (corresponding to \eqref{eq:FP}) 
\begin{equation}\label{eq:Lyapunov} 
     \gamma \dot\Sigma(t)=-K(t)\Sigma(t)-\Sigma(t) K(t)+2k_BT.
\end{equation}
In this case, the velocity field~\eqref{eq:vel-conservative} takes form
\begin{equation}\label{eq:vel-gaussian}
    v(t,x)=-\gamma^{-1} K(t)x +D\Sigma(t)^{-1}x,
\end{equation}
linear in $x$, since $\nabla \log\rho(t,x)= -\Sigma(t)^{-1}x.$

When the potential remains constant, with $K(t)=K_c$ symmetric and positive definite, the system reaches a steady-state distribution, which is Gaussian $\mathcal N(0,\Sigma_{ss})$ with the steady-state covariance $\Sigma_{ss}$ satisfying the algebraic Lyapunov equation
\begin{equation}\label{eq:ARE}
K_c\Sigma_{ss}+\Sigma_{ss} K_c=2k_BT.    
\end{equation}
The solution of \eqref{eq:ARE} is unique and can be expressed as
\begin{equation*}
\Sigma_{ss} = 2k_B\int_0^\infty e^{-\tau K_c} Te^{-\tau K_c}d\tau
 =L_{K_c}(2k_BT),   
\end{equation*}
where
\[
X \mapsto L_{A}(X) :=  \int_0^\infty   e^{-\tau {A}} Xe^{-\tau {A}'}d\tau,
\]
is a linear operator that depends on ${A}$. 

It is seen that the detailed balance condition ($J=0$) is special in that it requires that $K_c$ and $T$ commute. This holds when $K_c$ is diagonal, since $T$ is already diagonal. In this case, $\Sigma_{ss} = k_BTK_c^{-1}$ is also diagonal and results in zero probability current according to~\eqref{eq:vel-gaussian}; heat cannot transfer between the degrees of freedom.
When $K_c$ and $T$ do not commute, detailed balance breaks down and a non-vanishing probability current materializes leading to a non-equilibrium steady-state with non-vanishing heat transfer between the heat baths. 

We now depart from this steady-state analysis and focus on entropy production in a dynamic setting, for the special case of Gaussian distributions.

\subsection{Geometric decomposition of entropy production}
The  geometric decomposition of entropy production of Section \ref{sec:geometric}, specialized to the case of a Gaussian path of distributions, is as follows. Always assuming zero-mean, the path corresponds to a curve of covariance matrices $\{\Sigma(t):t\in[0,t_f]\}$, while the entropy production is given by
\begin{align*}
\int_0^{t_f}\dot S_{\rm tot}dt=\gamma \int_0^{t_f}\trace[V' T^{-1}V\Sigma]dt.
\end{align*}
Here, following \eqref{eq:v-decomp}, $v=Vx=-\left(\gamma^{-1} K -D\Sigma^{-1}\right)x$ admits the orthogonal decomposition
\begin{equation}
    \label{eq:decvfG}
Vx=\underbrace{TAx}_{T\nabla \phi}+\underbrace{\Omega\Sigma^{-1}x}_{\chi},
\end{equation}
where $A$ is a symmetric matrix and $\Omega$ is skew-symmetric, possibly time-varying.
Accordingly, thanks to the orthogonality condition $\trace[A\Omega]=0$, the entropy production decomposes into two parts
\begin{equation}\label{eq:decomp-gaussian}
    S_{\rm ex}+S_{\rm hk}=\gamma\hspace{-3pt}\int_0^{t_f}\hspace{-2pt}\trace[ATA\Sigma]dt+\gamma\hspace{-1pt}\int_0^{t_f}\hspace{-2pt}\trace[\Omega'{T^{-1}}\Omega\Sigma^{-1}]dt,
\end{equation}
in agreement with \eqref{eq:decomp}.

Echoing the development in Section \ref{sec:constrained}, for the special case of Gaussian densities and quadratic potential, we decompose the circulation and potential into their steady-state and dynamical components.
At steady-state (where $A=0$), the decomposition \eqref{eq:decvfG} implies that
$$
D\Sigma^{-1}-\gamma^{-1}K_{ss}=\Omega_{ss}\Sigma^{-1}.
$$
Using the fact that the right hand side of  $\gamma^{-1}K_{ss}=D\Sigma^{-1}-\Omega_{ss}\Sigma^{-1}$ must be symmetric (since $K_{ss}$ is), we obtain
\[
\Omega_{ss}\Sigma^{-1}+\Sigma^{-1}\Omega_{ss}= D\Sigma^{-1}-\Sigma^{-1}D,
\]
and hence,
\begin{align*}
    \Omega_{ss}&=L_{\Sigma^{-1}}(D\Sigma^{-1}-\Sigma^{-1}D).
\end{align*}
Similarly, since 
$\Omega_{ss}=D-\gamma^{-1} K_{ss}\Sigma$ must be skew-symmetric, we obtain
\begin{align*}
    K_{ss}&=2k_BL_{\Sigma}(T).
\end{align*}
When not at steady-state, $A\neq 0$ in \eqref{eq:decvfG} introduces an extra term leading to
\begin{subequations}
   \begin{align*}
    \Omega&= \Omega_{ss}-L_{\Sigma^{-1}}(TA-AT),\\
    K&=K_{ss}-\gamma L_{\Sigma}(TA\Sigma+\Sigma A T),
\end{align*} 
\end{subequations}
similarly to \eqref{eq:circulation-decomp}.

\subsection{Directions of vanishing housekeeping entropy production}

We now explain the content of
Proposition \ref{prop:0hk}, that characterizes directions with vanishing housekeeping entropy production, as it pertains to the Gaussian case.

In light of \eqref{eq:decvfG}, for zero housekeeping entropy production ($\chi=0$), the velocity field must be of the form $v=TAx$, with $A$ symmetric. At the same time, $v=-(\gamma^{-1}K-k_BT\Sigma^{-1})x$. Thus, $K$ which is symmetric, must be such that $T^{-1}K$ is also symmetric. (This argumentation recapitulates the reasoning that leads to $\nabla U$ being in the range of $\Pi_{\rho T^{-1}}$.) It follows that $K$ must be diagonal. In conclusion, the system is steered in a direction with vanishing housekeeping entropy production if and only if $K$ is diagonal.

We can further identify choices of $K$ that, besides ensuring $\dot S_{\rm hk}=0$, minimize 
excess entropy production \emph{rate} echoing Proposition \ref{prop:min-ex-0hk}. 
In the present case where $\nabla U=K x$ with $K$ diagonal, 
\begin{align*}
    \dot S_{\rm ex} &=\gamma \|D\nabla \log \rho + \gamma^{-1}\nabla U\|^2_{\rho T^{-1}}\\
    &=\gamma^{-1} \|k_BT\Sigma^{-1}x - Kx\|^2_{\rho T^{-1}}\\
    &=\gamma^{-1} \trace(k_B^2 \Sigma^{-1}T - 2k_BK + KT^{-1}K\Sigma).
\end{align*}
This is minimized for $K_{ii}=k_BT_i(\Sigma_{ii})^{-1},$ in agreement with \eqref{eq:psi0hkex}.

 \subsection{Direction of least entropy production rate}
We can readily obtain the potential function (i.e., the gain $K(t)$) that minimizes the  entropy production \emph{rate}. Indeed, with $U(x)=x' K x/2$ and Gaussian distribution~\eqref{eq:G-pdf}, equation  \eqref{eq:tomatch1} translates into
\begin{equation*}\label{eq:tomatch2} 
    T^{-1} K \Sigma+  \Sigma K T^{-1} =2k_B I,
\end{equation*}
or, equivalently, 
\begin{equation*}\label{eq:tomatch2a}
K\Sigma T +T \Sigma K=2k_BT^2,
\end{equation*}
with unique solution 
\[
K = 2 k_B L_{T\Sigma}(T^2).
\]

\subsection{Minimal entropy production between end-points}

Next we specialize the first-order optimality condition \eqref{eq:FONC} for {\em transition between end-point states} to Gaussian states and transition path. We adopt the ansatz that the Lagrange multiplier is of the form
$$
\lambda(t,x)=\frac12 x'\Lambda(t)x+c(t).
$$
The optimal $K,\Sigma$ and $\Lambda$ satisfy
\begin{subequations}
\begin{align*}
\gamma\dot\Sigma=&-K\Sigma-\Sigma K+2k_BT,\\
\gamma\dot\Lambda=&\Lambda K+K\Lambda+2KT^{-1}K+2k_B^2\Sigma^{-1}T\Sigma^{-1},\\
\gamma \dot c=&-2k_B^2\,\trace(T\Sigma^{-1})-k_B\,\trace[T\Lambda],\\
K =& -\frac{1}{2} L_{T\Sigma }(T\Lambda\Sigma T+T\Sigma\Lambda T),
\end{align*}
\end{subequations}
translating \eqref{eq:FONC} to the quadratic actuation case.
This is a set of coupled algebraic-differential equations with two-point boundary conditions, that can be solved numerically (e.g., via a shooting method).

\subsection{Minimal entropy production between Gaussian product states}

We now focus on minimizing entropy production while transitioning between two Gaussian states with mutually independent degrees of freedom, i.e., with diagonal covariances. In analogy with Section \ref{sec:prod-states}, this is equivalent to solving the unconstrained problem \eqref{eq:dynamicW2}, subject to a quadratic potential with end-points $\rho_0=\mathcal N(0,\Sigma_0)$ and $\rho_f=\mathcal N(0,\Sigma_f)$ with $\Sigma_0$ and $\Sigma_f$ diagonal. 
To see this, 
note that the optimal solution to \eqref{eq:dynamicW2} 
with Gaussian end-points is given by the Gaussian interpolation \eqref{eq:displacement-interpol}, where $\Sigma(t)$ remains diagonal at all times, as $\Sigma_0$ and $\Sigma_f$ are diagonal (recall that $M$ in \eqref{eq:displacement-interpolb} is diagonal). Consequently, $K(t)$ in \eqref{eq:Lyapunov} must also be diagonal. Since $\Sigma(t)$ and $K(t)$ are diagonal, the matrix $(-\gamma^{-1}K(t) + D\Sigma(t)^{-1})$ that defines the velocity field is also diagonal. Hence, the velocity field has no circulation ($\Omega=0$) and thus, the housekeeping entropy production vanishes. Evidently, this is due to the fact that $T$, $K$ and $\Sigma$ are ``aligned.''

The above considerations tell us very little about optimal trajectories that start or end at non-equilibrium steady-states. Next, we provide insight to such a scenario for a two-dimensional setting.

{
\section{A two-dimensional case}\label{sec:2D}

We now consider that $n=2$, where we may picture a particle with two degrees of freedom subject to respective stochastic excitations that correspond to temperatures $T_1>T_2$.  
This system has attracted considerable attention in works that have focused on quantifying heat transfer and torque produced in stationary states~\cite{BGyrator2007first,Bgyrator2013ciliberto,BGyrator2013dotsenko,BGyrator2017electrical,BGyrator2017experimental}.
The goal of the present section is to characterize entropy production, resulting from heat transfer as well as the dynamics in a non-stationary setting, and to determine time-varying control that precisely minimizes this entropy production.

\subsection{Explicit expressions of $S_{\rm ex}$ and $S_{\rm hk}$}
Starting from \eqref{eq:decomp-gaussian}, the total entropy production can be expressed as
\begin{align*}
\gamma\int_0^{t_f}\trace[ATA\Sigma]dt+\gamma\int_0^{t_f}\trace[\Sigma^{1/2}T^{-1}\Sigma^{1/2}\tilde \Omega\tilde \Omega'],
\end{align*}
where $\tilde\Omega=\Sigma^{-1/2}\Omega\Sigma^{-1/2}$ is still a skew-symmetric matrix.
Since we have defined our decomposition so that
$$
V\Sigma=-\gamma^{-1}K\Sigma+\gamma^{-1}k_BT=TA\Sigma+\Omega,
$$
we write
$$
\gamma V\Sigma+\gamma\Sigma V'=-K\Sigma-\Sigma K+2k_BT=\gamma TA\Sigma +\gamma \Sigma A T.
$$
Using this expression, the Lyapunov equation \eqref{eq:Lyapunov} becomes
\begin{align}\nonumber
\dot\Sigma&=TA\Sigma+\Sigma AT, \mbox{ or,}\\
\label{eq:lyap-A}
\dot{\tilde\Sigma}&=\tilde A\tilde \Sigma+\tilde \Sigma \tilde A,
\end{align}
for $\tilde A:=T^{1/2}AT^{1/2}$ and $\tilde \Sigma:=T^{-1/2}\Sigma T^{-1/2}$.
The excess entropy production can now be written as
\begin{align*}
    S_{\rm ex}&=\frac{\gamma}{2}\int_0^{t_f}\trace[\dot{\Sigma} A]dt=\frac{\gamma}{2}\int_0^{t_f}\trace[\dot{\tilde\Sigma}\tilde A]dt\\
    &=\frac{\gamma}{2}\int_0^{t_f}\trace[ L_{\tilde \Sigma}(\dot{\tilde \Sigma})\dot{\tilde\Sigma}]dt.
\end{align*}
 To obtain the second equality above we have used the identity $\tilde A=  L_{\tilde \Sigma}(\dot{\tilde\Sigma})$ as the solution to \eqref{eq:lyap-A}.
 
To simplify calculations we introduce the parametrization
\begin{equation}
\tilde\Sigma(r,\theta)=R\Big(\hspace{-3pt}-\frac{\theta}{2}\Big)\sigma^2(r)R\Big(\frac{\theta}{2}\Big),
\end{equation}
where
\begin{equation*}
    R(\vartheta)\!=\!\left[\hspace{-3pt}\begin{array}{cc}\!\cos(\vartheta)\!&\!\sin(\vartheta)\!\\\!-\sin(\vartheta)\!&\!\cos(\vartheta)\!\end{array}\hspace{-1pt}\right]\mbox{ and }
   \sigma^2(r) =\frac{l_c^2}{\sqrt{T_1T_2}}\left[\hspace{-1pt}\begin{array}{cc}\!e^r\!&\!0\!\\\!0&e^{-r\!}\end{array}\hspace{-3pt}\right],
\end{equation*}
are matrices, orthogonal and diagonal, respectively, and
where $l_c=\sqrt[4]{\det(\Sigma(t))}$\label{page:charlength} is a (constant) {\em characteristic length} of the system.
With this parametrization, one can explicitly express the excess part of entropy production as (see \cite{EnergyHarvestingAnisotropic2021} for similar computations and more details)
\begin{equation}\label{eq:S_ex}
S_{\rm ex}=k_B\tau\int_0^{t_f}\left(\cosh(r)\dot r^2+\sinh(r)\tanh(r)\dot\theta^2\right)dt,
\end{equation}
where $\tau=\gamma l_c^2/(2 k_B\sqrt{T_1T_2})$ is a {\em characteristic time} constant in that it is the average time that a Brownian motion with intensity $\sqrt{2\gamma^{-1}k_B\sqrt{T_1T_2}}$ needs to traverse a distance $l_c$. 
Note that \eqref{eq:S_ex} is quadratic in the velocities $(\dot r,\dot \theta)$ and therefore vanishes as $t_f\to\infty$. It is precisely the weighted Wasserstein action integral, expressed in terms of $r$ and $\theta$.

Let us look back at the housekeeping term and 
define $\omega$ through 
$$
\tilde\Omega=\omega\hat \Omega \mbox{ with } \hat\Omega=\left[\begin{array}{cc}0&-1\\1&0\end{array}\hspace{-3pt}\right].
$$
Thereby, $\tilde\Omega\tilde \Omega'=\omega^2I$ and the housekeeping entropy production can be written as
\begin{align*}
  S_{\rm hk}&=
\gamma\int_0^{t_f}\omega^2\trace[\Sigma^{1/2}T^{-1}\Sigma^{1/2}]dt=\gamma\int_0^{t_f}\omega^2\trace[\tilde\Sigma]dt\\&=\frac{2\gamma l_c^2}{\sqrt{T_1T_2}}\int_0^{t_f}\omega^2\cosh(r)dt,  
\end{align*}
where $\omega$ is to be determined so that
\[
K=- \gamma TA-\omega\gamma\Sigma^{1/2}\hat\Omega \Sigma^{-1/2}+k_BT\Sigma^{-1}
\]
is symmetric. 
Imposing $\trace[K\hat\Omega]=0$, which is equivalent to $K$ being symmetric, we obtain 
$$
\omega=\frac{\Delta T}{2}\frac{\big(\dot r+\tau^{-1}\sinh(r)\big)\sin \theta +\dot\theta\tanh(r)\cos\theta}{\bar T\cosh(r)+\Delta T\sinh(r)\cos\theta},
$$
where $\Delta T:=T_1-T_2$ and $\bar T=T_1+T_2$. Hence,
\begin{align}\label{eq:hkintermsof}
S_{\rm hk}&= k_B\tau\Delta T^2\int_0^{t_f}\frac{1}{\cosh(r)}\left(\frac{\sigma(r,\dot r,\theta, \dot \theta)}{\bar T+\Delta T\tanh(r)\cos\theta}\right)^2dt,
\end{align}
with
$$
\sigma(r,\dot r,\theta, \dot \theta)=\big(\dot r+\tau^{-1}\sinh(r)\big)\sin \theta +\dot\theta\tanh(r)\cos\theta.
$$
\black
\subsection{Directions of vanishing housekeeping entropy production}\label{sec:Shk}
Having derived explicit expressions of $S_{\rm ex}$ and $S_{\rm hk}$, equations \eqref{eq:S_ex} and \eqref{eq:hkintermsof},
it is illuminating to consider points and trajectories on which $S_{\rm hk}$ vanishes. Clearly this happens when $\theta(t)=0$, $\theta(t)=\pi$, or $r(t)=0$.
This should come as no surprise since these parameters render the covariance $\Sigma(t)$ diagonal. However, this is not the only trajectory for which $S_{\rm hk}=0$. 

\begin{prop}
    Trajectories corresponding to vanishing housekeeping entropy $S_{\rm hk}$ are given by solutions to the system of equations
    \begin{subequations}\label{eq:vanishingShk}
       \begin{align}\label{eq:vanishingShk1}
    \dot \theta&=\tan(\theta) u(t)\\\label{eq:vanishingShk2}
    \dot r&=- \tau^{-1}\sinh(r)-\tanh (r)u(t),
\end{align} 
    \end{subequations}
for any choice of function $u(t)$. 
\end{prop}
\begin{proof}
Housekeeping entropy vanishes iff $\sigma(r,\dot r, \theta, \dot\theta)=0$. Separating the variables $r$ and $\theta$ concludes
$$
\frac{\dot\theta}{\tan \theta}=-\tau^{-1}\cosh(r)-\frac{\dot r}{\tanh(r)}.
$$
Setting both sides equal to an arbitrary function of time $u(t)$ leads to \eqref{eq:vanishingShk}.
\end{proof}

It is interesting to study solutions of \eqref{eq:vanishingShk}, and thereby
flows that maintain $S_{\rm hk}=0$. We explore this next.

\begin{figure}
    \centering
\includegraphics[width=0.495\textwidth]{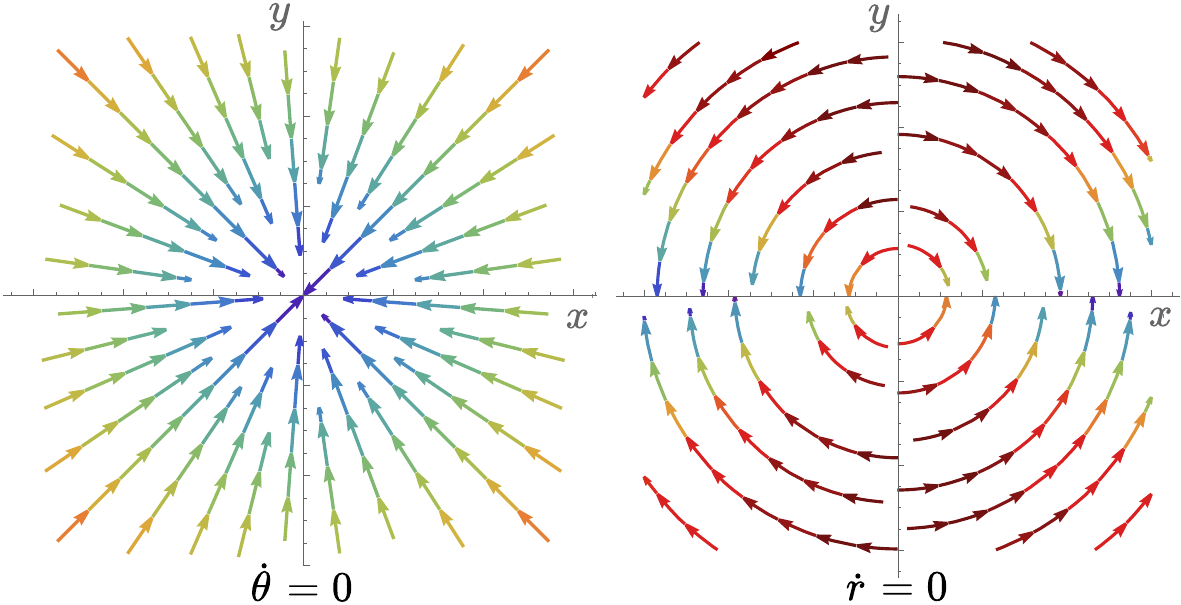}
    \caption{Trajectories of vanishing $S_{\rm hk}$ in the $(x,y)$ plane, where $x=r\cos(\theta)$ and $y=r\sin(\theta)$, and $\tau=1$ in \eqref{eq:vanishingShk2}.
    Left subplot: $u(t)$ is chosen such that $\dot\theta=0$. Right subplot: $u(t)$ is chosen such that $\dot r=0$.}
    \label{fig:zero-hk}
\end{figure}

We first note that the choice $u(t)=\frac{\kappa}{\tan(\theta_0+\kappa t)}$, with $\theta_0=\theta(0)$ and $\kappa\geq 0$, reduces \eqref{eq:vanishingShk1} to $\dot\theta=\kappa$, while \eqref{eq:vanishingShk2} becomes
\begin{equation}\label{eq:rdot-hk}
    \dot r=-\tau^{-1}\sinh(r)-\frac{\kappa}{\tan(\theta_0+\kappa t)}\tanh(r) \quad r(0)=r_{0}.
\end{equation}
This equation has a unique solution as long as $\theta_0+\kappa t\in (0,\pi)\,{\rm mod} \, \pi$. 
If we fix $\dot\theta=0$, this is, $u(t)=0$, then $\dot r=-\tau^{-1}\sinh(r)$ is always negative (but for when $r=0$). Therefore, these trajectories always point towards the equilibrium point at $r=0$.
Similarly, one can choose $u(t)$ to keep $\dot r$ constant. If $\dot r$ is kept at $0$, then $\dot\theta=-\tau^{-1}\cosh(r)\tan(\theta)$. These trajectories point towards either the $\theta=0$ or $\theta=\pi$ equilibrium points depending on the initial state.
The two sets of trajectories (with $\dot\theta=0$ and $\dot r=0$) are displayed in Figure~\ref{fig:zero-hk} for different initial conditions. 

Thus, interestingly, from any initial state one can find trajectories with vanishing housekeeping entropy production that steer the system to any of the possible equilibrium states ($\theta\in\{0,\pi\}$, or $r=0$).
On the other hand, steering between states (not necessarily equilibrium states) while maintaining $S_{\rm hk}=0$ impinges upon the controllability of the control affine system
\eqref{eq:vanishingShk}
viewing $u(t)$ as a control input.
It is clear that both $\theta=\{0,\pi\}$ and $r=0$ constitute obstructions to controllability of \eqref{eq:vanishingShk}, since the right hand sides of \eqref{eq:vanishingShk1} and \eqref{eq:vanishingShk2} vanish, respectively. Moreover, from any state $(r,\theta)$, with $\theta\notin\{0,\pi\}$ and $r\neq0$, a flow that maintains $S_{\rm hk}=0$ can be selected within a cone of $\pi$ radians. Specifically, accessible directions $\tan^{-1}(r\dot \theta/\dot r)$ from $(r,\theta)$ are within 
the interval $[\alpha,\alpha+\pi]$, for $$\alpha=\tan^{-1}\left(-\frac{r\tan(\theta)}{\tanh (r)}\right).$$

In light of \eqref{eq:vanishingShk}, directions and therefore trajectories with vanishing housekeeping entropy production can not have arbitrarily small velocity fields; this can be traced to the fact that to eliminate circulation, the dynamical component in the decomposition \eqref{eq:circulation-decomp} needs to cancel the steady-state component. This leads unavoidably to positive excess entropy production ($S_{\rm ex}$).

While maintaining $\dot S_{\rm hk}=0$, we seek tangent directions $(\dot r,\dot\theta)$ that also minimize excess entropy, i.e.,
\begin{align*}
    \min_u\  k_B\tau\int_0^{t_f}\left(\cosh(r)\dot r^2+\sinh(r)\tanh(r)\dot\theta^2\right)dt,
\end{align*}
with $\dot r$ and $\dot \theta$ as in \eqref{eq:vanishingShk}. This leads to the optimal choice of $u(t)$,
$$
u^*(t)=-\tau^{-1}\frac{\cosh(r)}{1+(\tan\theta)^2}.
$$
Corresponding trajectories are displayed in the left subplot of Figure \ref{fig:sdotmin}, for a choice of parameters, and converge to a $\theta\in\{0,\pi\}$ equilibrium state.

\subsection{Direction of least entropy production rate}\label{sec:Srate}

We characterize the direction of minimal entropy production \emph{rate} in the following proposition.

\begin{prop}\label{prop:prop6} For any given $(r,\theta)\in [0,\infty)\times [0,2\pi)$, the directions $(\dot r,\dot \theta)$ that minimize the  entropy production rate $\dot S$
are given by
\begin{subequations}\label{eq:minErateVels}
    \begin{align}
    \dot r&=-\tau^{-1}h(r,\theta)\sinh(r)\sin(\theta),\\
      \dot \theta&=-\tau^{-1}h(r,\theta)\cosh(r)\cos(\theta),
\end{align}
\end{subequations}
where $$h(r,\theta)=\frac{(\tanh(r))^2\sin(\theta)}{(\frac{\bar T}{\Delta T}+\cos(\theta)\tanh(r))^2(\sinh(r))^2+(\tanh(r))^2}.$$
\end{prop}

\vspace*{5pt}

\begin{proof}
From
\eqref{eq:S_ex} and \eqref{eq:hkintermsof}, the entropy rates for $\dot S_{\rm ex}$ and $\dot S_{\rm hk}$ are
\begin{align*}
    \dot S_{\rm ex}&=k_B\tau\left(\cosh(r)\dot r^2+\sinh(r)\tanh(r)\dot\theta^2\right),\\
    \dot S_{\rm hk}&=k_B\tau\frac{\Delta T^2 }{\cosh(r)}\left(\frac{\sigma(r,\dot r,\theta, \dot \theta)}{\bar T+\Delta T\tanh(r)\cos\theta}\right)^2.
\end{align*}
    Then, letting $\xi=\left[
         \dot r, 
         \dot \theta
   \right]'$, $\dot S_{\rm tot}=\dot S_{\rm ex}+\dot S_{\rm hk}$ can be written as
    $$
(k_B\tau)^{-1}\dot S_{\rm tot}=\xi'A\xi+b'\xi+c,
    $$
    where 
   \begin{align*}
        c&=\tau^{-2}\Delta T^2\tanh(r)\sinh(r)(\sin\theta)^2/d(r,\theta),
    \\
    b&=  \frac{2\tau^{-1}\Delta T^2\tanh(r)\sin(\theta)}{d(r,\theta)}\left[\begin{array}{c}
   \sin\theta\\ \tanh(r)\cos(\theta)
    \end{array}\right]
   \end{align*}
   and
\[
A=\left[\hspace*{-1pt}
\begin{array}{cc} 
          \cosh(r)+\frac{\Delta T^2 (\sin(\theta))^2}{\cosh(r)d(r,\theta)} & \frac{\Delta T^2 \sin(\theta)\cos(\theta)\tanh(r)}{\cosh(r)d(r,\theta)}
         \\ 
   \frac{\Delta T^2 \sin(\theta)\cos(\theta)\tanh(r)}{\cosh(r)d(r,\theta)} & \hspace{-3pt}\frac{(\sinh r)^2}{\cosh(r)}+\frac{\Delta T^2 (\cos(\theta)\tanh(r))^2}{\cosh(r)d(r,\theta)}
\end{array}\hspace{-1pt}\right],
\]
with $d(r,\theta)=(\bar T+\Delta T\tanh(r)\cos\theta)^2$.
  Since $A$ is positive definite for all $r$ and $\theta$, we obtain that $\xi=-A^{-1}b'/2$ minimizes $\dot S_{\rm tot}$ over all possible $\xi$.
\end{proof}

 Optimal trajectories for different initial conditions are drawn in the right subplot in Figure \ref{fig:sdotmin}. It is worth noting that these streamlines are similar \emph{in form} to those to the left, that correspond to trajectories that minimize excess entropy production while vanishing housekeeping entropy production. 


\begin{figure}
    \centering
\includegraphics[width=0.49\textwidth,trim={0.05cm 0cm 0cm 0cm},clip]{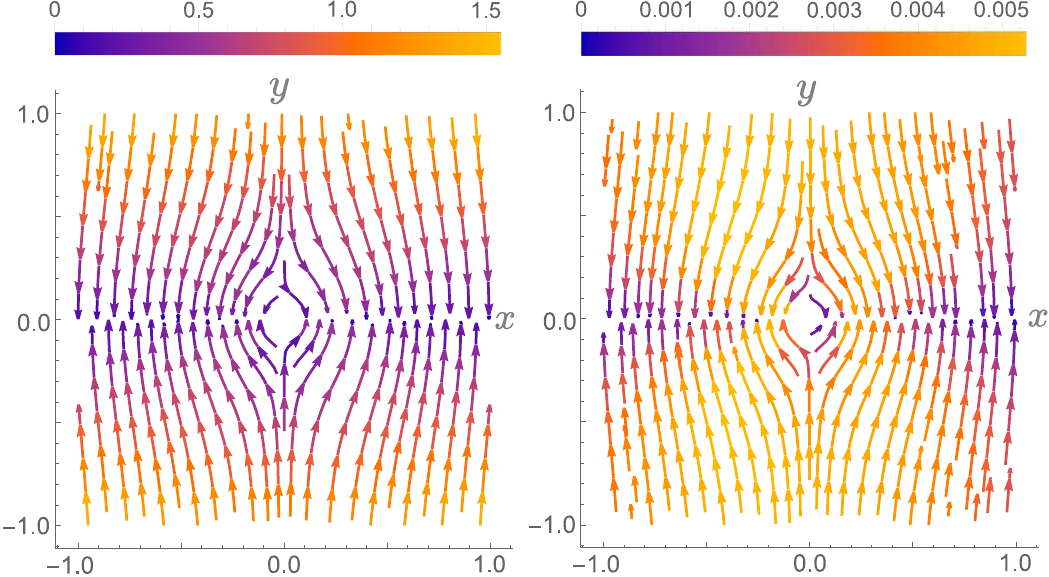}
    \caption{Left: Trajectories  that minimize excess entropy production while vanishing housekeeping entropy production. Right: Trajectories that
    minimize entropy production rate $\dot S_{\rm tot}$. All trajectories are displayed in coordinates $x=r\cos(\theta)$, $y=r\sin(\theta)$ for the following choice of parameters: 
    $\tau=1=\bar T$ and $\Delta T=0.1$.
  }
    \label{fig:sdotmin}
\end{figure}

\subsection{Minimal entropy production close to equilibrium}
We now return to the problem of minimizing total entropy production over a path that joins two (possibly non-equilibrium) end-points in finite time.
To gain insight as to the nature of optimal trajectories,
we consider staying close to the equilibrium that corresponds to $r=0$.


To this end, we let $r(t)=\epsilon \hat r (t)$, $\hat x=\hat r \cos\theta$ and $\hat y=\hat r \sin\theta$, and expand the expression for $S$ in terms of $\epsilon>0$, assumed small, to obtain
    \begin{align*}
S_{\rm tot}&=\epsilon^2k_B\tau\int_0^{t_f} \bigg(\dot{\hat r}^2+\hat r^2\dot \theta^2+\Big(\frac{\Delta T}{\bar T}\Big)^2 \big((\dot{\hat r}+\tau^{-1}\hat r)\sin\theta\\
&\mbox{\phantom{.}}\hspace{132pt}+\dot \theta \cos\theta \hat r\big)^2
   \bigg)dt+O(\epsilon^3)\\&= \underbrace{\epsilon^2k_B\tau\int_0^{t_f} \Big(\dot{\hat x}^2+\dot{\hat y}^2+\Big(\frac{\Delta T}{\bar T}\Big)^2 \big(\dot{\hat y}+\tau^{-1}\hat y\big)^2\Big)dt}_{S_\epsilon(\hat x,\hat y)}+O(\epsilon^3).
\end{align*}
Thus, the entropy production up to second order in $\epsilon$, $S_\epsilon(\hat x,\hat y)$, is as specified above.

\begin{prop}
The trajectory that minimizes entropy production between two states close to the equilibrium at $r=0$, up to second order,
 \begin{align}\label{eq:problemrsmall}
(\hat x^*,\hat y^*)={\rm argmin}\{S_\epsilon(\hat x,\hat y)\mid &\;\hat x(0)=\hat x_0,\ \hat y(0)=\hat y_0,\\
&\; \hat x(t_f)=\hat x_{t_f},\ \hat y(t_f)=\hat y_{t_f}
\},\nonumber
\end{align}
is of the form
\begin{subequations}\label{eq:optimalxy}
  \begin{align}
    \hat x^*(t)&=\hat x_0+\frac{t}{t_f}(\hat x_{t_f}-\hat x_0),\\
    \hat y^*(t)&= c_+ e^{t/\hat\tau }+c_-e^{-t/\hat\tau }.
\end{align}  
\end{subequations}
 where
$$ 
     \hat\tau=\tau\big(1+\big(\tfrac{\bar T}{\Delta T}\big)^2\big)^{1/2}  \mbox{ and }
     c_\pm=\frac{\hat y_0 -\hat y_{t_f}e^{\pm t_f/\hat\tau }}{1-e^{\pm 2t_f/\hat\tau}}.
$$
\end{prop}

\vspace*{5pt}

\begin{proof}
The Euler-Lagrange equations for minimizing $S_\epsilon$ take the form
   \begin{align}
       \ddot{\hat x}&=0,\quad  \hat x(0)=\hat x_0,\ \hat x(t_f)=\hat x_{t_f}, \\
       \ddot{\hat y}&=\frac{1}{\hat\tau^2} \hat y,\quad  \hat y(0)=\hat y_0,\ \hat y(t_f)=\hat y_{t_f}.
   \end{align}
   Solving these equations and imposing the boundary conditions we obtain the sought result.
\end{proof}


  It is also insightful to consider fixing the starting state at $(\hat x_0,\hat y_0)$, near the equilibrium at $r=0$ as before, and consider the trajectory departing from this state with terminal time  $t_f\to\infty$ and the final state  unconstrained. The optimal solution is then given by $\hat x(t)=\hat x_0$ and $\hat y(t)=\hat y_0e^{- t/\hat\tau}$, as any other $\hat y(t)$ would lead to infinite entropy production. Therefore, as long as $r$ is small enough, trajectories minimizing entropy production over an unbounded time interval end up at one of the $\theta\in\{0,\pi\}$ equilibrium states (as opposed to the one corresponding to $r=0$).
  
  \begin{remark}
 It is interesting to observe that trajectories minimizing total entropy production over an infinite time window, for $r$ small enough, share some resemblance with those of vanishing housekeeping entropy production while minimizing excess entropy rate, as well as those minimizing total entropy production rate, portrayed in the left and right subplots of Figure \ref{fig:sdotmin}, respectively. Indeed, they approach equilibrium for $\theta\in\{0,\pi\}$ almost vertically and have $\tau$ as a natural time constant. $\Box$
  \end{remark}

\subsection{Entropy minimizing cycles}

\begin{figure}[t]
    \centering
\includegraphics[width=0.5\textwidth,trim={0.325cm 0cm 0cm 0.15cm},clip]{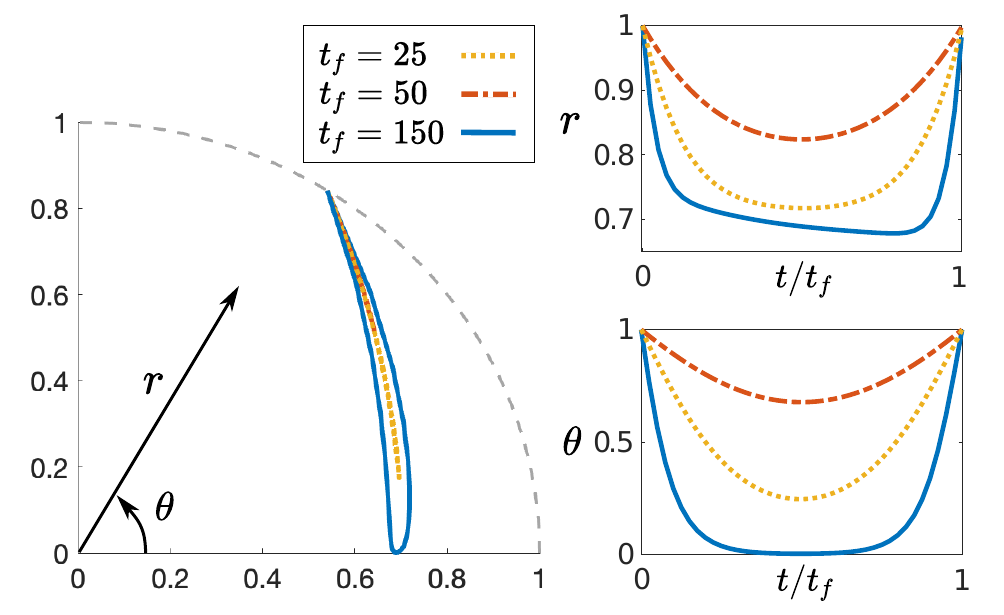}
    \caption{Entropy minimizing cycles starting and ending at $(r,\theta)=(1,1)$, with final times $t_f=10,\,25,\,50$ and $150$ seconds, respectively. The rest of the parameters are set to $1$, but for $\Delta T=0.1$.}
    \label{fig:minE}
\end{figure}



In this last section we turn to cycles that generate minimal entropy production. The expressions we obtain did not allow for closed-form solutions, and hence we have resorted to numerically computing optimal trajectories. Valuable insights are gained in that we observe a natural tendency of trajectories to gravitate towards an equilibrium state, as much as time allows, before returning to their starting point. This is in stark contrast to the isotropic case with a single heat bath, in which optimal trajectories are trivially constant.

\black
 We selected as the starting and ending point $(r_0,\theta_0)=(1,1)$, and computed closed trajectories, portrayed in Figure~\ref{fig:minE}, of different periods. The natural tendency is to gravitate towards an equilibrium state where $S_{\rm hk}=0$, interestingly, one that corresponds to $\theta=0$. We have seen the same tendency in the analysis of  small $r$, where entropy minimizing trajectories over an infinite time window converge to an equilibrium at $y=0 \Leftrightarrow \theta=0$; this is apparent in Figure \ref{fig:thetarsmall}.

 \begin{figure}
    \centering
\includegraphics[width=0.365\textwidth,trim={0cm 0.1cm 0cm 0.5cm},clip]{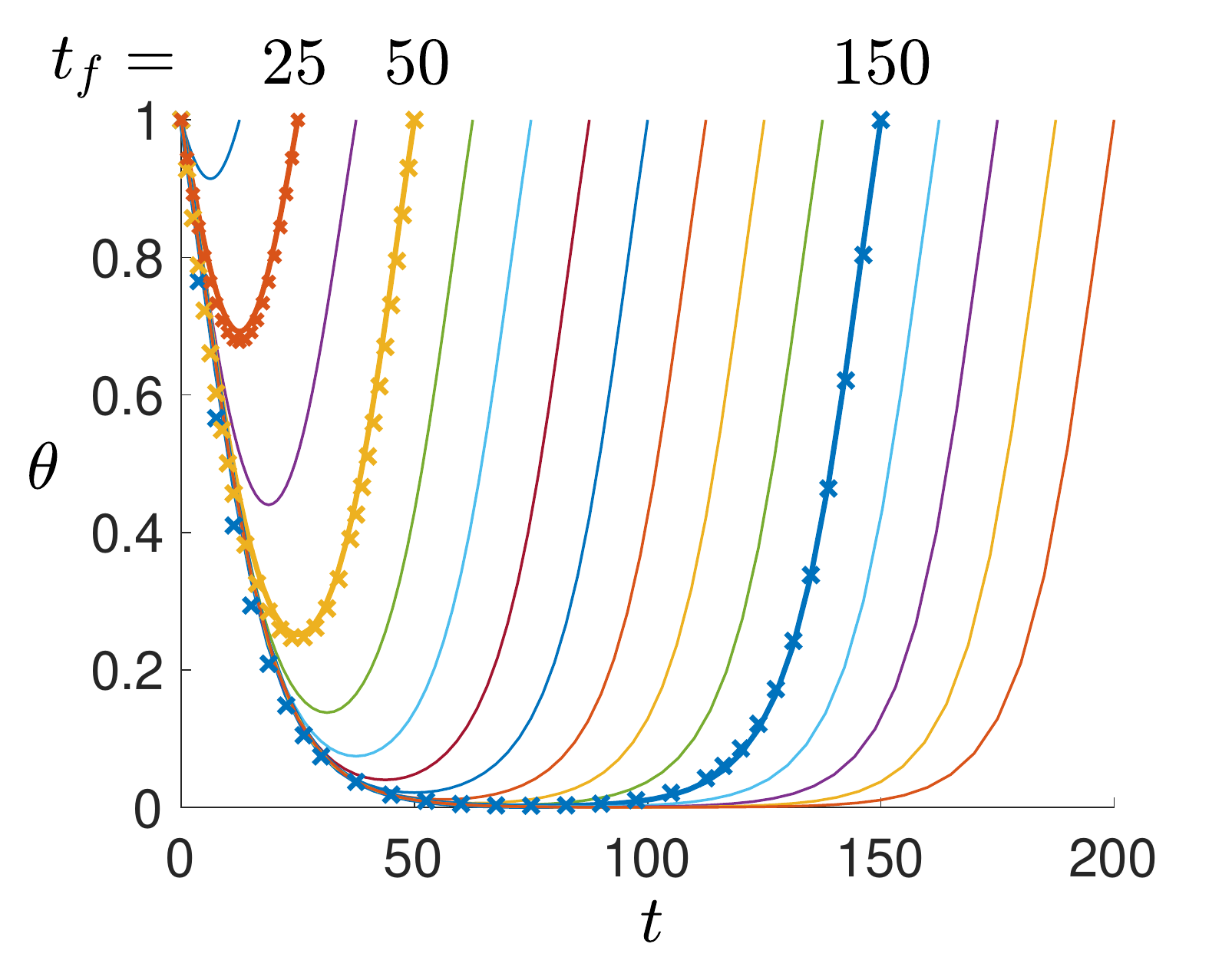}
    \caption{Solid lines represent $\theta$ component of entropy minimizing cycles in the limit of vanishing $r$ for different final times $t_f$.  Crosses denote the $\theta$ component of entropy minimizing cycles obtained numerically (not requireing $r$ small). The rest of the parameters are set to $1$, but for $\Delta T=0.1$.} 
    \label{fig:thetarsmall}
\end{figure}

 Figure \ref{fig:thetarsmall} compares the analytic solution when minimizing entropy production for $r$ vanishingly small \eqref{eq:problemrsmall}, to the numerical solution in Figure \ref{fig:minE} that is computed for $r$ large. At the outset, there is no apparent reason why these should compare. Yet,
 it is observed that the $\theta$ component of trajectories, {\em whether $r$ is small or not}, follows a similar path. 
 To see this, we drew in Figure \ref{fig:thetarsmall} the $\theta$ component of the analytic solution \eqref{eq:optimalxy} as a function of time (continuous curve), this is $\theta(t)={\tan^{-1}}\big(\frac{\hat y^*(t)}{\hat x^*(t)}\big)$ with $\hat x^*$ and $\hat y^*$ as in \eqref{eq:optimalxy}. A very good agreement with the numerical solution 
 is readily seen (marked with $\times$'s).  The $r$-components cannot be compared and thus respective plots are omitted.
 
 An additional reason for focusing on the $\theta$ component of trajectories is that this component reveals an apparent natural time constant $\hat\tau$ that dictates an optimal balance between $S_{\rm hk}$ and $S_{\rm ex}$, as the trajectories approach a suitable point of equilibrium at $\theta=0$. Indeed, when the period is sufficiently large, trajectories spend any ``extra'' time allocated near the equilibrium, as seen in Figure~\ref{fig:thetarsmall}.
 Interestingly, within the time-window that the trajectory stays near equilibrium (near $\theta=0$), assuming $t_f$ is sufficiently large, $r$ linearly decreases as seen in Figure~\ref{fig:minE}. Apparently this helps in reducing the housekeeping cost during the return leg of the trajectory back to $(r,\theta)=1$.}

 A final important point is that entropy minimizing cycles are work consuming. Specifically, to steer the thermodynamic state along these closed trajectories, the controlling input (actuating potential) needs to supply work to the system. This is due to the fact that these trajectories are traversed clockwise (as can be seen from the asymmetry in the $r$ plot), which based on
 previous work~\cite{EnergyHarvestingAnisotropic2021} indicates  work consumption.  
We do not further expand on this point since
  it is tangential to the present work.
  However, we would like to underscore the importance
  of controlling the thermodynamic system for positive work output, balancing the work produced with total entropy generated. This optimal control problem remains open at present.
\black

\black
\section{Conclusions}

In these pages we sought to understand
how to limit 
entropy production that is inherent to stochastic systems with anisotropic thermal excitation.
We highlighted the necessity of non-conservative forcing for stalling entropy production,   
identified sources that contribute to entropy production, and characterized 
control actions that minimize them. In doing this,  the structure inherited from a naturally weighted inner product has taken a central role shaping optimal protocols and trajectories.
We explicitly worked out a two-dimensional case so as to illustrate the problem's intrinsic quirks, such as the existence of zero-housekeeping entropy producing trajectories and the propensity towards equilibrium of entropy minimizing cycles.

Several theoretical issues remain, such as the construction of general optimal entropy-minimizing controls, the robustness of control protocols to uncertainty in the constituents and the environment and, most importantly, the study of trade-offs between work extraction and entropy production.
In our previous works \cite{EnergyHarvestingAnisotropic2021,miangolarra2022geometry} we focused on steering a thermodynamic system, similarly subject to anisotropic temperatures, over a cycle that maximizes work extraction. In light of the present results, it is worth considering maximizing work output subject to a bound on the entropy production over a cycle. Such a bound is natural, especially in biological engines, where sources that sustain chemical gradients are taxed by their entropy production.
Optimal control of thermodynamic systems for the purpose of trading off entropy for work may prove essential in the understanding of biological mechanisms that enable life.

\appendices

\section*{Acknowledgments}
The research was supported in part by the AFOSR under grant FA9550-20-1-0029, and ARO under W911NF-22-1-0292. O.M.M was supported by ”la Caixa” Foundation (ID 100010434) with code LCF/BQ/AA20/11820047.

\bibliography{arXiv}
\bibliographystyle{IEEEtran}

\section*{Appendix} 
\label{app}

Fluctuation theorems quantify the arrow of time by comparing the likelihood of observing a certain response of the system to that of  observing the time-reversed response. In other words, they can be viewed as performing a Neyman-Pearson test to validate a hypothesis on the time-arrow. It turns out that the likelihood relates to entropy production that highlights the ``strength'' and correct direction of the time-arrow.  Typically, earlier statements of such theorems, assumed isotropic temperature environment. Herein we present a similar derivation for a fluctuation result that holds in anisotropic environments.
\black

Consider an overdamped system, as in \eqref{eq:Langevin}, but driven instead by a force field $F(\lambda(t),x)=-\nabla U(\lambda(t),x)+f(\lambda(t),x)$,
 \begin{equation}\nonumber
dX_t=\gamma^{-1}
F(\lambda(t),X_t)dt+ \sqrt{2D}dB_t,\ X_0\sim \rho(0,\cdot),
 \end{equation}
where $\lambda(t)$ is a time-dependent control parameter, $t\in[0,t_f]$, and $X_t$ is distributed according to $\rho(t,x)$.
Denote by
$P^\lambda_{X}$ 
the probability measure induced
by the process $X$  on the space of continuous paths $x=\{x_t: t\in[0,t_f]\}$.
We define the time-reversal of trajectories and protocols,  $\tilde{x}_t := x_{t_f - t}$ and $\tilde{\lambda}(t) := \lambda(t_f  - t)$, as well as
a suitably time-reversed process driven by the same Brownian motion,
\begin{equation}\nonumber
d\tilde X_t=\gamma^{-1}
F(\tilde \lambda(t),\tilde X_t)dt+ \sqrt{2D}dB_t , \ \tilde X_0\sim\tilde\rho(0,\cdot),
 \end{equation}
 where $\tilde\rho(0,\cdot)=\rho(t_f,\cdot)$.
Our goal is to
 compare the probability that $X=x$ with that of $\tilde X= \tilde{x}$. 
 
Denote by $P_{Z}$ the stationary Wiener measure induced by $ \ud Z_t  = \sqrt{ 2 D}\ud B_t$, with $Z_0$ distributed according to the Lebesgue measure\footnote{$P_Z$ is an unbounded positive measure on paths (hence, not a probability measure) and $Z_t$ is referred to as the reversible Wiener process \cite{leonard2014some}.}. 
The Radon-Nikodym derivative of $P^\lambda_{X}$ with respect to $P_{Z}$  (reference measure) is
\begin{align*}
\frac{\ud P^\lambda_{X}}{\ud P_{Z}} = \exp\Bigg[\sum_{i=1}^n\Big(&\frac{1}{2 k_B T_i} \int_0^{t_f} F_i(\lambda(t),X_t) (\ud X_t)_i \\&- \frac{1}{4\gamma k_B T_i}  \int_0^{t_f}  | F_i(\lambda(t),X_t)|^2 \ud t\Big)\Bigg]\rho(0,\cdot), 
\end{align*}
see \cite[Section 3.5]{karatzas1991brownian}; it represents a probability density with respect to the reference measure. 
A similar expression can be written for the ``backward'' law $P^{\tilde\lambda}_{\tilde X}$.

We now wish to evaluate the ratio
\begin{equation}\label{eq:ratio}
\frac{dP^\lambda_{X}(x)}{d P^{\tilde\lambda}_{\tilde X}(\tilde x)}=\left(\frac{\ud P^\lambda_{X}}{\ud P_{Z}}(x)\right) \text{\LARGE $/$}
\Bigg(\frac{\ud  P^{\tilde\lambda}_{\tilde X}}{\ud P_{Z}}(\tilde x)\Bigg),
\end{equation}
this is, the ratio between the probability of the forward path $x$ under the forward law, and the probability of the backward path $\tilde x$ under the ``backward'' law.

Since $\int_0^{t_f}  | F_i(\lambda(t),X_t)|^2 \ud t$ is an ordinary integral, its value is unaffected by time reversal of the integrand, and therefore, setting $X=x$ and $\tilde X=\tilde x$ gives that
$$ \int_0^{t_f}  | F_i(\lambda(t),x_t)|^2 \ud t=  \int_0^{t_f}  | F_i(\tilde\lambda(t),\tilde x_t)|^2 \ud t
$$
for all $i$.
On the other hand, when considering stochastic integrals care must be taken.
Here, $\int_0^{t_f}F_i(\lambda(t), X_t)  (\ud  X_t)_i$ is approximated in mean-square by sums
\[
\sum_j F_i(\lambda(t_j), x_{t_j})(x_{t_{j+1}}-x_{t_j}),
\]
and likewise, $\int_0^{t_f}F_i(\tilde \lambda(t), \tilde X_t)  (\ud  \tilde X_t)_i$ can be approximated at the time-reversed trajectories by
\[
\sum_j F_i(\lambda(t_{j+1}),x_{t_{j+1}})(x_{t_j}-x_{t_{j+1}}).
\]
Their difference shows that
\begin{align*}
 &  \int_0^{t_f} F_i(\lambda(t),X_t) (\ud X_t)_i\mid_x -\int_0^{t_f}F_i(\tilde\lambda(t),\tilde X_t)  (\ud \tilde X_t)_i\mid_{\tilde x}\\&=  2\int_0^{t_f} F_i(\lambda(t),X_t) \circ (\ud X_t)_i\mid_x =-2q_i,  
\end{align*}
 where we have used the definition of $q_i$ in~\eqref{eq:dq}.
 
Therefore, the ratio in \eqref{eq:ratio} becomes
 \[
\frac{dP^\lambda_{X}(x)}{d P^{\tilde\lambda}_{\tilde X}(\tilde x)}=\exp\left(-\sum_{i=1}^n \frac{q_i}{k_B T_i} \right)\frac{\rho(0,x_0)}{\rho(t_f,x_{t_f})}.
\]
Noting that  the total entropy difference between end points is given by 
 $\Delta s_{\rm tot}=-k_B\big(\log\big(\rho({t_f},x_{t_f})\big)-\log\big(\rho({0},x_{0})\big)\big)-\sum_{i=1}^n \frac{q_i}{T_i} $ we obtain
 \begin{equation}\label{eq:likelihood}
\frac{dP^\lambda_{X}(x)}{d P^{\tilde\lambda}_{\tilde X}(\tilde x)}= \exp\left(\frac{\Delta s_{\rm tot}}{k_B}\right).
 \end{equation}
Thus, a large entropy difference signifies that a trajectory is (exponentially) more likely to occur than its time-reversed counterpart. Taking the expected value of the inverse in \eqref{eq:likelihood} with respect to $dP^\lambda_X$, we finally obtain \eqref{eq:FT}.

\end{document}